\documentclass[a4paper,12pt]{amsart}
\usepackage[left=1.2in, right=1.2in, , bottom = 1.2in, top = 1.2in]{geometry}
\usepackage[utf8]{inputenc}
\usepackage[all]{xy}
\usepackage{amssymb}
\usepackage{amsmath}
\usepackage{csquotes}
\usepackage{subfig}
\usepackage{enumerate}
\usepackage{graphicx}
\usepackage{enumerate}
\usepackage{amsaddr}
\usepackage{color}
\usepackage{mwe}
\usepackage{subfig}
\usepackage{appendix}

\usepackage{soul}
\usepackage{xcolor}
\newcommand\independent{\protect\mathpalette{\protect\independenT}{\perp}}
\def\independenT#1#2{\mathrel{\rlap{$#1#2$}\mkern2mu{#1#2}}}

\usepackage{xr}
\makeatletter
\newcommand*{\addFileDependency}[1]{
  \typeout{(#1)}
  \@addtofilelist{#1}
  \IfFileExists{#1}{}{\typeout{No file #1.}}
}
\makeatother

\usepackage{tikz}
\usetikzlibrary{arrows.meta,shapes}
\usepackage{setspace}

\title[]{Separable Effects for Causal Inference in the Presence of Competing Events}
\author{Mats J. Stensrud$^{1,2}$, Jessica G. Young$^3$, Vanessa Didelez$^{4,5}$,James M. Robins$^{1,6}$,  Miguel A. Hern\'{a}n$^{1,6,7}$} \address{ $^1$ Department of Epidemiology, Harvard T. H. Chan School of Public Health, USA \\
$^2$Department of Biostatistics, University of Oslo, Norway\\
$^3$ Department of Population Medicine,
Harvard Medical School and Harvard
Pilgrim Health Care Institute, USA \\
$^4$ Leibniz Institute for Prevention Research and Epidemiology – BIPS, Germany \\
$^5$ Faculty of Mathematics / Computer Science,
University of Bremen, Germany \\
$^6$ Department of Biostatistics, Harvard T. H. Chan School of Public Health, USA  \\
$^7$ Harvard-MIT Division of Health Sciences and Technology, USA  \\}
\date{\today}

\MakeOuterQuote{"}\EnableQuotes
\DeclareUnicodeCharacter{00A0}{ }

\begin{document}
\begin{abstract}
In time-to-event settings, the presence of competing events complicates the definition of causal effects. Here
we propose the new \textsl{separable effects} to study the causal effect of a treatment on an event of interest. The separable \textsl{direct} effect is the treatment effect on the event of interest not mediated by its effect on the competing event. The separable \textsl{indirect} effect is the treatment effect on the event of interest only through its effect on the competing event. Similar to Robins and Richardson's extended graphical approach for mediation analysis, the separable effects can only be identified under the assumption that the treatment can be decomposed into two distinct components that exert their effects through distinct causal pathways. Unlike existing definitions of causal effects in the presence of competing events, our estimands do not require cross-world contrasts or hypothetical interventions to prevent death. As an illustration, we apply our approach to a randomized clinical trial on estrogen therapy in individuals with prostate cancer. 
\end{abstract}

\maketitle

\section{Introduction}
A competing event is any event that makes it impossible for the event of interest to occur. For example, consider a randomized trial to estimate the effect of a new treatment on the 3-year risk of prostate cancer in which 1000 individuals with prostate cancer were assigned to the treatment and 1000 to placebo. All participants adhered to the protocol and remained under follow-up. After 3 years, 100 individuals in the treatment arm and 200 in the placebo arm died of prostate cancer. Also, 150 individuals in the treatment arm and 50 in the placebo arm died of other causes (e.g.,\ cardiovascular disease). Death from cardiovascular disease is a competing event for death from prostate cancer: individuals who die of
cardiovascular disease cannot subsequently die of prostate cancer. When competing events are present, several causal estimands may be considered to define the causal effect of treatment on a time-to-event outcome \cite{young2018causal}.
 
Consider first the total treatment effect \cite{young2018causal} defined by the contrast of the cumulative incidence (risk) \cite{prentice1978analysis, andersen2012competing} of the event of interest under different treatment values. In our example, the total treatment effect on death from prostate cancer is the contrast of the cumulative incidence of death from prostate cancer under treatment, consistently estimated by $\frac{100}{1000}$, and under placebo, consistently estimated by $\frac{200}{1000}$. Therefore, the estimate of the total treatment effect on the additive scale is $\frac{100}{1000}-\frac{200}{1000}=-0.1$, which indicates that
treatment reduced the risk of death from prostate cancer.
 
However, in our trial, the interpretation of the total treatment effect on the event of interest is difficult because the treatment also increased the risk of the competing event. The estimate of the total effect of treatment on the competing event is $\frac{150}{1000}-\frac{50}{1000}=0.1$ on the additive
scale. Thus, it is possible that the beneficial effect of treatment on death from prostate cancer is simply a consequence of the harmful effect of treatment on death from other causes: when more people die from other causes, fewer people can die from prostate cancer. Note that this problem of interpretation cannot be solved by considering contrasts of hazard functions, such as cause-specific and subdistribution hazards, because these estimands are defined conditional on a post-treatment event (survival) and therefore do not generally have a causal interpretation \cite{young2018causal, hernan2010hazards}.

One way to deal with this problem is to consider a second causal estimand on the risk scale: the (controlled) direct effect of treatment on the event of interest had competing events been eliminated. This estimand corresponds to defining the competing events as censoring events \cite{young2018causal}, and is sometimes denoted the marginal (net) distribution function. Unlike the total effect, identification of the controlled direct effect requires untestable assumptions even in an ideal randomized trial with perfect adherence and no loss to follow-up \cite{young2018causal}. Also, this causal estimand often introduces a new conceptual challenge: the direct effect is not sufficiently well-defined because there is no scientific agreement as to which hypothetical intervention, if any, would eliminate the competing events \cite{hernan2016does}. For example, in our prostate cancer trial, no intervention has ever been proposed that can prevent all deaths from causes other than prostate cancer. As a byproduct of the ill-defined intervention to prevent competing events, effect estimates cannot be empirically verified -- not even in principle -- in a randomized experiment.

A third causal estimand is the survivor average causal effect (SACE) \cite{robins1986new}, which is the total treatment effect (on the risk scale) in the principal stratum of patients who would never experience the competing event under either level of treatment \cite{young2018causal,robins1986new, frangakis2002principal}. Unlike the total effect, the presence of competing events is not a problem when interpreting the SACE, because the SACE is restricted to subjects who do not experience competing events. However, identification of the SACE requires strong untestable assumptions, e.g.\ about cross-world counterfactuals, even in a perfectly executed trial. Also, the SACE could never, even in principle, be confirmed in a real-world experiment as it will never be possible to observe the status of the competing event for the same individual under two different levels of treatment.

%\textcolor{red}{In the failure time literature, it has  been common to consider estimands on the hazard scale, in particular cause-specific and subdistribution hazards, but contrasts of hazards are even harder to interpret causally as they are defined conditional on a post-treatment event (survival) \cite{young2018causal}. }

The problems of the previous estimands can be overcome in settings in which the treatment exerts its effect on the event of interest and its effect on the competing event through different causal pathways. Here, we define the separable direct and indirect effects for settings with competing events. Like the controlled direct effect and the SACE, identification of separable effects relies on untestable assumptions even when the treatment is randomized. However, unlike the controlled direct effect and the SACE, separable effects do not require conceptual interventions on competing events or knowledge of cross-world counterfactuals; the separable effects are well-defined if we can articulate a hypothetical decomposition of the treatment into two components. Therefore, in principle, they may be verified in a future experiment. Our definitions of separable effects and conditions for identifiability follow from the work of Robins and Richardson \cite{robins2010alternative} and Didelez \cite{didelez2018defining} on mediation: the pure (natural) direct effects \cite{robins1992identifiability} are extensively used in mediation analyses, but they require untestable cross-world independence assumptions and are often difficult to interpret, for example, in survival settings. Robins and Richardson \cite{robins2010alternative} proposed an alternative causal estimand that overcomes these problems by considering a decomposed treatment: unlike the pure direct effects, the decomposed treatment effects can be identified under assumptions that are in principle empirically testable. Moreover, it was shown by Didelez \cite{didelez2018defining} that the decomposed treatment effects are sensible estimands in survival settings. 

%in mediation settings where the pure (natural) direct effect \cite{robins1992identifiability} is of substantial importance, Robins and Richardson \cite{robins2010alternative} argued that this estimand can be re-expressed as an effect of a decomposed treatment. The decomposed treatment effect can, unlike the pure direct effect, be identified under assumptions that are empirically testable. Didelez \cite{didelez2018defining} used this treatment decomposition to motivate mediation estimands in a survival setting. 

%Like the usual direct effect estimand, ours is defined as the direct effect of treatment on the event of interest that is not mediated through effects on the competing events. Unlike the usual direct effect estimand, ours explicitly articulates an hypothetical intervention to prevent the competing events, thus ameliorating interpretational concerns and allowing for the possibility that the effect estimate can be empirically verified in a randomized experiment. However, our causal estimand is only well-defined in settings in which treatment exerts independent effects on the event of interest and on the competing events. The definition of our estimand and its identifiability conditions follow from the work of Robins and Richardson\cite{robins2010alternative} and Didelez \cite{didelez2018defining}.

We have organized the paper as follows. In Section \ref{sec: data structure}, we describe the observed data structure. In Section \ref{sec: decomposition}, we present a conceptual treatment decomposition and provide explicit examples to fix ideas. In Section \ref{sec: estimand def}, we formulate the causal estimand and define the new separable effects. In Section \ref{sec: id separable effects}, we present conditions that allow for identifiability of the separable effects. In Section \ref{sec: estimation separable effects}, we give 3 different estimators for the separable effects that can be implemented with standard statistical models, and we use data from a randomized clinical trial to estimate a direct effect of estrogen therapy on prostate cancer mortality. In Section \ref{sec: discussion}, we provide a final discussion of the new estimands. 

\section{Observed data structure}
\label{sec: data structure}
We consider a study in which individuals are 
randomly assigned to a binary treatment $A\in \{0,1\}$ at baseline (e.g.\ $A=1$ if assigned to treatment and $A=0$ if assigned to placebo). Let $L \in \mathcal{L}$ denote a vector of individual pretreatment characteristics. For each of equally spaced discrete time intervals $k \in \{0,1,...,K+1\}$, let $Y_{k}$ and $D_{k}$ denote indicators of an event of interest and a competing event by interval $k$, respectively. In our example, $Y_{k}$ denotes death due to prostate cancer and $D_{k}$ death from other causes by interval $k$. We adopt the convention that $D_{k}$ is measured just before $Y_{k}$. If an individual experiences the competing event at time $k$ without a history of the event of interest $(D_{k}=1,Y_{k-1}=0)$, then all future values of the event of
interest are zero. We can approximate a continuous time setting by choosing time intervals that are arbitrary small. 

By definition, $D_{0}\equiv Y_{0}\equiv 0$, that is, no individual experiences any event during the initial interval. We use overbars to denote the history of a
random variable, such that $\bar{Y}_{k}=(Y_{1},Y_{2},...,Y_{k})$ is the history of the event of interest through interval $k$. Similarly, we use underbars to denote future values of a random variable, such that $\underbar{Y}_{k}=(Y_{k},Y_{k+1},...,Y_{K+1})$. We assume full adherence to the assigned treatment without loss of generality, and until Section \ref{sec: general ID}, no loss to follow-up.

\section{Decomposition of treatment effects}
\label{sec: decomposition}
Suppose that treatment $A$ can be conceptualized as having two binary components that act through different causal pathways: one component $A_{Y}$ that affects the event of interest $Y_{k}$ and one component $A_{D}$ that affects the competing event $D_{k}$. This hypothetical decomposition of $A$ can be formally described by the following conditions.

Suppose that $A$ and the two components $A_{Y}$ and $A_{D}$ are deterministically related in the observed data,
\begin{align}
    A\equiv A_{D}\equiv A_{Y},
    \label{assumption: Determinsm}
\end{align}
but we can conceive hypothetical interventions that set $A_{D}$ and $A_{Y}$ to different values. For $k \in \{0,...,K\}$, let $Y_{k+1}^{a}$ be an individual's indicator of the event of interest at time $k+1$ when, possibly contrary to fact, $A$ is set to the value $a\in \{0,1\}$. Similarly, let $Y_{k+1}^{a_Y,a_D}$ be this outcome when, possibly contrary to fact, $A_{Y}$ is set to $a_Y$ and $A_{D}$ is set to $a_D$, where $a_Y,a_D\in \{0,1\}$. We require that setting $A=a$ is equivalent to setting both $A_{Y}$ and $A_{D}$ to $a$, that is, 
\begin{align}
Y_{k+1}^{a_Y=a,a_D=a} & =Y_{k+1}^{a}, \nonumber \\
D_{k+1}^{a_Y=a,a_D=a} & =D_{k+1}^{a}, \text{ for } k \in \{0,K\}.
\label{eq: definition A=Ay=Ad}
\end{align}
%When we consider hypothetical interventions, such as in Section \ref{sec: statins and dementia}, our confidence that the analysis will be valid depends on the plausibility of \eqref{eq: definition A=Ay=Ad}.
The assumption that $A_{D}$ only exerts effects on $Y_{k+1}$ through its effect on $\overline{D}_{k+1}$ can be stated as 
\begin{align}
& Y_{k}^{a_Y,a_D=1}=D_{k+1}^{a_Y,a_D=1}=Y_{k}^{a_Y,a_D=0}=D_{k+1}^{a_Y,a_D=0}=0 \implies \nonumber \\
& \qquad Y_{k+1}^{a_Y,a_D=0}=Y_{k+1}^{a_Y,a_D=1}, \quad \text{ for } a_Y \in \{0,1\},  \label{eq: implication 1}
\end{align}
and, similarly, the assumption that $A_{Y}$ only exerts effects on $D_{k+1}$ through its effect on $\overline{Y}_k$ can be stated as 
\begin{align}
& Y_{k}^{a_Y=1,a_D}=D_{k}^{a_Y=1,a_D}=Y_{k}^{a_Y=0,a_D}=D_{k}^{a_Y=0,a_D} =0 \implies \nonumber \\
& \qquad D_{k+1}^{a_Y=1,a_D}=D_{k+1}^{a_Y=0,a_D}, \quad \text{ for } a_D \in \{0,1\}.  \label{eq: implication 2} % \label{eq: implication 2}.
\end{align}

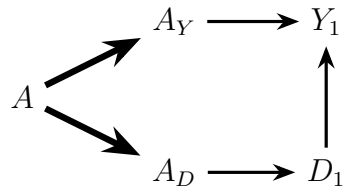
\begin{figure}
\centering
\begin{tikzpicture}
\begin{scope}[every node/.style={thick,draw=none}]
    \node (A) at (0,0) {$A$};
    \node (Ay) at (2,1) {$A_{Y}$};
	\node (Ad) at (2,-1) {$A_{D}$};
	\node (Y1) at (4,1) {$Y_1$};
    \node (D1) at (4,-1) {$D_1$};
\end{scope}

\begin{scope}[>={Stealth[black]},
              every node/.style={fill=white,circle},
              every edge/.style={draw=black,very thick}]
    \path [->] (A) edge[line width=0.85mm] (Ad);
    \path [->] (A) edge[line width=0.85mm] (Ay);
	\path [->] (Ad) edge (D1);
    \path [->] (Ay) edge (Y1);	
    \path [->] (D1) edge (Y1);
\end{scope}
\end{tikzpicture}
\caption{Directed acyclic graph for a trial with a randomized baseline treatment $A$, such that $A_{Y}$ and $A_{D}$ are deterministic functions (bold arrows) of $A$, competing event $D_1$ and event of interest $Y_1$ at time 1 of follow-up, with $D_1$ measured just before $Y_1$.}
\label{Figure:DAGsimple}
\end{figure}

The causal diagram in Figure \ref{Figure:DAGsimple} represents this decomposition in a setting with a single time point. The bold arrows represent the deterministic relation \eqref{assumption: Determinsm}. Our decomposition conditions do not preclude the existence of multiple forms of decompositions of $A$. However, every decomposition of $A$ into two distinct components must be justified by subject-matter knowledge. Let us consider two examples.

\subsection{Diethylstilbestrol and prostate cancer mortality}
\label{sec: DES part 1}
In our prostate cancer example, we assume that $A$ can be decomposed into a component $A_{Y}$ that directly affects death from prostate cancer and a component $A_{D}$ that directly affects death from other causes. Suppose that treatment $A=0$ is placebo and $A=1$ is diethylstilbestrol (DES), an estrogen which is thought to reduce mortality due to prostate cancer by suppressing testosterone production and to increase cardiovascular mortality through estrogen-induced synthesis of coagulation factors \cite{turo2014diethylstilboestrol}.

We could then consider a hypothetical treatment that has the same direct effect as DES on prostate cancer mortality, but lacks any effect effect on mortality from other causes; that is, the same effect as the $A_Y$ component of DES when the $A_D$ component is removed. Real-life treatments similar to such a hypothetical treatment are luteinizing hormone releasing hormone (LHRH) antagonists or orchidectomy
(castration), which can stop testosterone production but, unlike estrogen, do not increase cardiovascular risk.

Also, we could consider a hypothetical treatment that has the same direct effect as DES on mortality from other causes, but that lacks any effect on prostate cancer mortality; that is, the same effect as the $A_D$ component of DES when the $A_Y$ component is removed. In practice, a drug that contains not only DES but also testosterone may resemble this hypothetical treatment, as the additional testosterone component can nullify the testosterone suppression that is induced by DES.

\subsection{Statins and dementia}
\label{sec: statins and dementia}
Consider a study to quantify the effect of statins on dementia. Statins reduce cardiovascular mortality by lowering the cholesterol production in the liver. As dementia may develop due to microvascular events in the small cerebral arteries, lowering cholesterol may also reduce the risk of dementia. When studying the effect of statins on dementia, death will be a competing event.

Because statins appear to reduce mortality and dementia through the same mechanism, i.e., lowering the cholesterol levels in the blood, decomposing $A$ into the distinct components $A_{Y}$ and $A_{D}$ would be difficult. One possibility might be to leverage the distinct localization of the microvessels in the brain: we could bioengineer a cholesterol transporter, which is surgically implanted to shuttle cholesterol particles from the distal cerebral arteries directly to the large cerebral veins, circumventing the cerebral microvessels. That is, if $Y_k$ and $D_k$ denote dementia and death, respectively, then carriers of the transporter will have the $A_{Y}$ component of statins on dementia, but they will lack the $A_{D}$ component of statins on mortality. Robins and Richardson discussed the construction of plausible interventions in a mediation context, using nicotine in cigarettes as an example \cite[Section 5.2]{robins2010alternative}.

\subsection{Practical considerations}
\label{sec: practical considerations}

%The separable effects described in the next section are sufficiently well-defined causal estimands 
Whenever the decomposition of treatment $A$ into $A_{Y}$ and $A_{D}$ is possible in principle, regardless of whether it is possible in practice at this time in history, the effects of $A_Y$ and $A_D$ are well-defined. Therefore, in both examples above, we described well-defined effects even though the decomposition of treatment may be practically possible in the prostate cancer example but not in the statin example. 

However, caution is required when considering treatment decompositions that, as in the statins example, are possible in principle but not in practice. The problem is that practically impossible decompositions make it hard to evaluate the identifiability conditions for the effects of each component. As described in Section 5, the identification of the separable effects is based on the unverifiable condition that $A_{Y}$ and $A_{D}$ are treatment components actually operating in the data \cite{hernan2016does}, such that $A_{Y}$ has no direct effect on $D_k$ and that $A_{D}$ has no direct effect on $Y_k$. When relying on convoluted treatment decompositions, as in our statins example, we may be less confident that these conditions hold in the data. Of course, if these conditions are violated, our effect estimates may differ from those that would be obtained in a future experiment in which both components $A_{Y}$ and $A_{D}$ are randomly assigned.

On the other hand, a careful definition of treatment decomposition may help ground scientific conversations even if the decomposition is not yet possible. For example, it is debated whether statins have a protective effect on dementia \cite{power2015statins}. To clarify the notion of a 'protective effect' it would be helpful to consider a hypothetical trial in which subjects were randomly assigned to the cholesterol transporter or placebo. 

\section{Definition of separable effects}
\label{sec: estimand def}
We can now define the \emph{separable direct effects of treatment on the
event of interest} as the contrasts 
\begin{equation*}
%\mathbf{Separable direct effects 1:}\quad
\Pr (Y_{k+1}^{a_Y=1,a_D}=1)\text{ vs. }\Pr
(Y_{k+1}^{a_Y=0,a_D}=1)
\end{equation*}
for $a_D=1$ or $a_D=0$; that is, the effect of the component of treatment that affects the event of interest $A_Y$ when the component of treatment that affects the competing event $A_D$ is set at a constant value $a_D$.

Analogously, we can define the \emph{separable indirect effects of treatment on the event of interest} as the contrasts 
\begin{equation*}
%\mathbf{Separable direct effects 1:}\quad 
\Pr (Y_{k+1}^{a_Y,a_D=1}=1)\text{ vs. }\Pr
(Y_{k+1}^{a_Y,a_D=0}=1),
\end{equation*}%
for $a_Y=1$ and $a_Y=0$; that is, the effect of the component of treatment that affects the competing event $A_D$ when the component of treatment that affects the event of interest $A_Y$ is set at a constant value $a_Y$. In other words, the separable indirect effects are functions of the treatment component $A_D$ that affects the competing event $D_{k+1}$, and the separable indirect effects arise because the competing event makes it impossible for the event of interest to occur.

From  \eqref{eq: definition A=Ay=Ad} we find that the sum of separable direct and indirect effects (on the additive scale) equals the total effect,
\begin{align*}
& [\Pr (Y_{k+1}^{a_Y=1,a_D=1}=1)-\Pr (Y_{k+1}^{a_Y=0,a_D=1}=1)] \\
& +[\Pr (Y_{k+1}^{a_Y=0,a_D=1}=1)-\Pr (Y_{k+1}^{a_Y=0,a_D=0}=1)] \\
& =\Pr (Y_{k+1}^{a=1}=1)-\Pr (Y_{k+1}^{a=0}=1).
\end{align*}
To provide intuition about the magnitude of the separable effects, we describe 4 illustrative scenarios in Appendix \ref{sec app: intuition sep eff}.

\section{Identification of separable effects}
\label{sec: id separable effects}
\label{sec: identification main} The identification of the separable
effects requires the identification of
the quantities 
\begin{align}
\Pr (Y_{k+1}^{a_Y,a_D}=1),
\label{eq: estimand}
\end{align}%
where $a_Y,a_D \in \{0,1\}$. Identifying these quantities would be straightforward if each of the treatment components could be separately intervened upon, that is, if we could conduct a randomized experiment with 4 possible treatment arms defined by the 4 combinations of values of $A_{Y}$ and $A_{D}$. However, when using data from a study like that of Section \ref{sec: data structure}, in which only the treatment $A$ is randomized, we only observe 2 out of the 4 treatment arms in a hypothetical trial in which $A_{Y}$ and $A_{D}$ were randomized. As a result, we need additional untestable conditions to identify \eqref{eq: estimand}. This conceptualization of the treatment decomposition in terms of a 4-arm randomized experiment was originally proposed by James Robins during a presentation at the UK Causal Inference Conference in London, April 2016. Since then, Robins and others have often publicly discussed this conceptualization in the context of mediation analysis, which is isomorphic to the context with competing events discussed here.

\subsection{Identifiability conditions}
\label{sec: simple id conditions}

First, we need exchangeability conditional on the measured covariates $L$,
\begin{align*}
& \bar{Y}_{K+1}^{a},\bar{D}_{K+1}^{a}\mathpalette{\protect\independenT}{\perp}A \mid L \text{ for all } a,
\end{align*}%
where time $K+1$ is the end of the study. This exchangeability condition is expected to hold when $A\equiv A_{Y} \equiv A_{D}$ is randomized. 

Second, consistency, such that if $A=a$, then 
\begin{align*}
&Y_{k+1}^{a}=Y_{k+1} \\
& D_{k+1}^{a}=D_{k+1},
\end{align*}
for $a\in\{0,1\}$ at all times $k\in\{0,\ldots,K\}$. If any subject has data history consistent with the intervention under a counterfactual scenario, then the consistency assumption ensures that the observed outcome is equal to the counterfactual outcome.

Third, positivity such that % for all $l \in \mathcal{L}$,
\begin{align}
 & \Pr(L=l)>0\implies   \notag \\
 & \quad \Pr (A=a\mid  L=l)>0 \text{ for } a\in\{0,1\}, %  \text{ and }  l \in \mathcal{L}, 
\label{eq: positivity of A} \\ 
  & \Pr(D_{k+1}=Y_{k}=0,L=l) > 0  \implies \nonumber\\ 
 & \quad  \Pr(A=a|D_{k+1}=Y_{k}=0,L=l)>0 \text{ for } a\in\{0,1\}\text{ and }k\in\{0,\ldots,K\}.  %\label{eq: positivity of A part 2}
%& \quad \Pr (A=a\mid  L)>0\text{ w.p.1} \text{ for } a\in\{0,1\},
\label{eq: positivity of A part 2}
\end{align}%
where \eqref{eq: positivity of A} is the usual positivity condition under interventions on $A$ and \eqref{eq: positivity of A part 2} ensures that among those event-free through each follow-up time, there exist individuals with $A=1$ and individuals with $A=0$. However, our estimand is based on hypothetical intervention on both $A_{Y}$ and $A_{D}$, and our positivity conditions do not ensure the stricter condition that
\begin{align*}
%& f( L)>0\implies   \notag \\
 & \Pr(L=l)>0\implies   \notag \\
& \quad \Pr (A_Y=a_Y, A_D=a_D \mid L=l)>0 \text{ for } a_Y,a_D \in\{0,1\},% \text{ and }  l \in \mathcal{L}  ,\text{ w.p.1}
%\label{eq: positivity of A}
\end{align*}%
which, indeed, will be violated when $a_Y \neq a_D$ in our setting where $A \equiv A_{Y} \equiv A_{D}$.  

To allow for identifiability under our positivity condition in \eqref{eq: positivity of A}, we introduce two conditions that are related to conditions described by Didelez in a mediation setting \cite{didelez2018defining}.

\subsection*{Dismissible component condition 1}
\begin{align*}
\mathbf{\Delta1:}& \Pr (Y_{k+1}^{a_Y,a_D=1}=1\mid Y_{k}^{a_Y,a_D=1}=0,D_{k+1}^{a_Y,a_D=1}=0, L = l) \\
& =\Pr (Y_{k+1}^{a_Y,a_D=0}=1\mid Y_{k}^{a_Y,a_D=0}=0,D_{k+1}^{a_Y,a_D=0}=0, L = l),
\end{align*}%
at all times $k \in \{0,..., K\}$. That is, the counterfactual (discrete-time) hazards of the event of interest are equal under all values of $A_{D}$.

\subsection*{Dismissible component condition 2}
\begin{align*}
\mathbf{\Delta2:}& \Pr (D_{k+1}^{a_Y=1,a_D}=1\mid Y_{k}^{a_Y=1,a_D}=0,D_{k}^{a_Y=1,a_D}=0, L = l) \\
& =\Pr (D_{k+1}^{a_Y=0,a_D}=1\mid Y_{k}^{a_Y=0,a_D}=0,D_{k}^{a_Y=0,a_D}=0, L = l),
\end{align*}%
at all times $k \in \{0,..., K\}$. That is, the counterfactual (discrete-time) hazard functions of the competing event are equal under all values of $A_{Y}$. The dismissible component conditions are analogous to identification conditions from Shpitser \cite{shpitser2013counterfactual} on path specific effects.

By considering a hypothetical trial in which both $A_Y$ and $A_D$ are randomized, we can define conditional independencies that imply the dismissible component conditions, and these conditional independencies can be read off of causal DAGs directly, see Appendix \ref{sec: DCC as independences} for details. 

\begin{figure}
\centering
\begin{tikzpicture}
\begin{scope}[every node/.style={thick,draw=none}]
    \node (A) at (0,0) {$A$};
    \node (Ay) at (2,1) {$A_{Y}$};
	\node (Ad) at (2,-1) {$A_{D}$};
	\node (Y1) at (4,1) {$Y_1$};
    \node (D1) at (4,-1) {$D_1$};
    \node (Y2) at (6,1) {$Y_2$};
    \node (D2) at (6,-1) {$D_2$};
    \node (U) at (3,3) {$U_{YD}$};
   % \node (D3) at (8,-1) {$D(3)$};
\end{scope}

\begin{scope}[>={Stealth[black]},
              every node/.style={fill=white,circle},
              every edge/.style={draw=black,very thick}]
    \path [->] (A) edge[line width=0.85mm] (Ad);
    \path [->] (A) edge[line width=0.85mm] (Ay);
	\path [->] (Ad) edge (D1);
    \path [->] (Ad) edge[bend right] (D2);
	\path [->] (Ay) edge[bend left] (Y2);
    \path [->] (Ay) edge (Y1);	
    \path [->] (U) edge (Y1);
    \path [->] (U) edge (D1);
    \path [->] (Y1) edge (D2);
    \path [->] (Y1) edge (Y2);
    \path [->] (D1) edge (D2);
    \path [->] (D1) edge (Y1);
    \path [->] (D2) edge (Y2);
\end{scope}
\end{tikzpicture}
\caption{Extension of the causal directed acyclic graph in Figure 1 which includes an unmeasured common cause $U_{YD}$  which violates conditions $\Delta$1 and $\Delta$2.}
\label{Figure:DAGviolationSimple}
\end{figure}
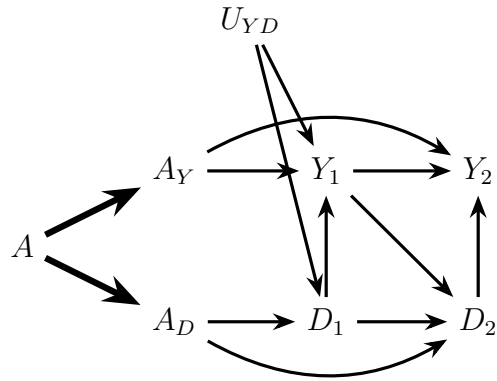

The dismissible component conditions ensure that we can adjust for common causes of $D_{k}$ and $Y_{k'}$ for all $k,k'\in \{1,...,K+1\}$. In particular, an unmeasured common cause of $D_1$ and $Y_{1}$, such as $U_{YD}$ in Figure \ref{Figure:DAGviolationSimple}, violates $\Delta$1 and $\Delta$2. In our prostate cancer example, suppose that smoking is a common cause of death from prostate cancer ($Y_k$) and death from other causes ($D_k$). Then, if smoking is an unmeasured variable (such as $U_{YD}$ in Figure \ref{Figure:DAGviolationSimple}), the dismissible component conditions will be violated.

%due to conditioning on a collider
However, the presence of unmeasured causes $U_Y$ of $Y_k$ and unmeasured causes $U_D$ of $D_k$, as shown in Figure \ref{Figure:DAGfrailty}, does not violate $\Delta$1 and $\Delta$2 (see Appendix \ref{sec: exploring 1 and 2} for details); it just implies that contrasts of the hazard terms in \eqref{eq: identification L} cannot be causally interpreted \cite{young2018causal, hernan2010hazards, stensrud2017exploring}, which is analogous to the mediation setting in Didelez \cite[Figure 6]{didelez2018defining}. For this reason, we have defined our causal estimands as contrasts of risks rather than as contrasts of hazards. Furthermore, adjusting for a measured common cause of $Y_k$ and $D_k$, such as $L$ in Figure \ref{Figure:DAGconfounder1}, allows identification under $\Delta$1 and $\Delta$2. In subsequent figures we have omitted the variables $U_Y$ and $U_D$ to avoid clutter, but our results are valid in the presence of $U_Y$ and $U_D$. We have also omitted an arrow from $L$ to $A$, but this arrow would not invalidate our results. Furthermore, we have intentionally omitted arrows from $D_k$ to $Y_{s}$ for $k<s$, as these arrows are redundant in our setting where the competing event is a terminating event that precludes the event of interest at all subsequent times. Finally, note that if the dismissible component conditions hold on a coarser scale (say, daily), then they will in general also hold on a finer scale (say, hourly), but the reverse is not true. This is analogous to any setting where measurements of time-varying covariates are needed to identify causal effects.

The dismissible component conditions are not empirically verifiable in a trial in which the entire treatment $A$, but neither of its components $A_{Y}$ and $A_{D}$, is intervened upon. However, both conditions could be tested in a trial in which $A_{Y}$ and $A_{D}$ were randomly assigned. 

\subsection{Identification formula}
Under the identifiability conditions in Section \ref{sec: simple id conditions}, we identify $\Pr
(Y_{k+1}^{a_Y,a_D}=1)$ from the following g-functional \cite{robins1986new} of the observed data described in Section \ref{sec: data structure},
\begin{align}
& \sum_{l}  \Big[ \sum_{s=0}^{k} \Pr(Y_{s+1}=1 \mid D_{s+1}= Y_{s}=0,A = a_Y, L=l) \nonumber \\ 
  &  \prod_{j=0}^{s}  \big[ \Pr(D_{j+1}=0 \mid D_{j}= Y_{j}=0,A = a_D, L=l)  \nonumber \\
      &  \times \Pr(Y_{j}=0 \mid D_{j}= Y_{j-1}=0,A = a_Y, L=l) \big] \nonumber \Big] \Pr(L=l),  \notag \\
\label{eq: identification L}
\end{align}%
%which can be shown by considering a hypothetical trial where $A_Y$ and $A_D$ are randomized, see Appendix \ref{sec: proof of identifiability},
%where we have assumed that $L$ is discrete (we can generalize to a continuous $L$ using product integrals). 
see Appendix \ref{sec: proof of identifiability} for proof. 

%Importantly, we must measure $L$, i.e.\ all common causes of the event of interest and the competing event, to identify the separable effects by \eqref{eq: identification L}, even when $A$ is randomized.

%\textcolor{red}{Notice that in settings where $a_Y=a_D$, it immediately follows from law of total probability that \eqref{eq: identification L} equals $\Pr(Y_{s+1}=1)$. When $a_Y \neq a_D$, \eqref{eq: identification L} only differs from $\Pr(Y_{s+1}=1)$ due to the second line, where we in \eqref{eq: identification L} condition on $A=a_D$, not $A=a_Y$. Thus, when the equality $\Pr(D_{k+1}=0 \mid D_{k}= Y_{k}=0,A = 1, L=l) = \Pr(D_{k+1}=0 \mid D_{k}= Y_{k}=0,A = 0, L=l)$ holds for all $k \in \{0,K\}$, then $\Pr(Y_{k+1}^{a_Y,a_D=1}=1) = \Pr(Y_{k+1}^{a_Y,a_D=0}=1) = \Pr(Y_{k+1}^{a=a_Y}=1)$}.

\subsection{Intuition on the identification formula \eqref{eq: identification L} and falsifiability of the separable effects.}
Identification formula \eqref{eq: identification L} can be intuitively motivated as follows: consider an experiment $G$ in which both $A_Y$ and $A_D$ are randomly assigned such that $\Pr(A_Y=a_Y,A_D=a_D) > 0$ for all $a_D,a_Y \in \{0,1\}$. In the experiment $G$, $\Pr(Y^{a_Y,a_D}_{k+1}=0) = \Pr(Y_{k+1}=1 \mid A_Y = a_Y, A_D = a_D)$ by randomization. By the laws of probability $\Pr(Y_{k+1}=1 \mid A_Y = a_Y, A_D = a_D)$ can in turn be re-expressed as
\begin{align}
& \sum_{l}  \Big[ \sum_{s=0}^{k} \Pr(Y_{s+1}=1 \mid D_{s+1}= Y_{s}=0,A_Y = a_Y,A_D = a_D, L=l) \nonumber \\ 
  &  \prod_{j=0}^{s}  \big[ \Pr(D_{j+1}=0 \mid D_{j}= Y_{j}=0,A_Y = a_Y,A_D = a_D, L=l)  \nonumber \\
      &  \times \Pr(Y_{j}=0 \mid D_{j}= Y_{j-1}=0,A_Y = a_Y,A_D = a_D, L=l) \big] \nonumber \Big] \Pr(L=l).  \nonumber \\
      \label{eq: outcome under ay and ad}
\end{align}%
Formula \eqref{eq: identification L} can be obtained by applying the dismissible component conditions to the terms in \eqref{eq: outcome under ay and ad}. These additional conditions are needed for identification in our current study because, unlike in $G$, only $A$ was randomized in our current study and not the separate components $A_Y$ and $A_D$.  If the experiment $G$ is actually conducted in the future, then the separable effect estimates obtained from \eqref{eq: identification L} in our current study can be confirmed by comparing them to estimates of $\Pr(Y_{k+1}=0 \mid A_Y = a_Y, A_D = a_D)$ from $G$ \cite{robins2010alternative}. 
%by comparing $\Pr(Y_{k+1}=0 \mid A_Y = a_Y, A_D = A_D)$ to estimates of the functional \eqref{eq: identification L} in $G$; that is, the dismissible component conditions are testable in principle

Note that \eqref{eq: identification L} can also be read off of a Single World Intervention Graph (SWIG) \cite{richardson2013single} that satisfies the dismissible component conditions, as suggested in Figure \ref{fig: simple SWIG}, illustrating that the separable effects are single-world quantities that are empirically testable in principle. This is in contrast to alternative approaches from mediation analysis that require additional, untestable cross-world independence assumptions \cite{robins2010alternative}.

%Also, note that if $A_Y \independent D_{k+1} \mid G$ in $G$, then $\Pr(Y^{a_Y,a_D}_{k+1}=1) = \Pr(Y^{a=a_Y}_{k+1}=1)$, which implies that the separable direct effects are 0 for all $a_D$. 

\subsection{Separable effects in the presence of censoring}
\label{sec: general ID}
We consider a subject to be censored at time $k+1$ if the subject remained under follow-up and was event-free until $k$, but we have no information about the subject's events at $k+1$ or later \cite{young2018causal}. That is, censoring is a type of event that does not make it impossible for the event of interest to occur and we assume that censoring can in principle be prevented \cite{young2018causal}. When the censoring is independent of future counterfactual events given $L$, as illustrated in Figure \ref{fig: censoring}, we can identify the separable effects from
\begin{align}
& \sum_{l}  \Big[ \sum_{s=0}^{k} \Pr(Y_{s+1}=1 \mid D_{s+1}= Y_{s}=\bar{C}_{s+1}=0,A = a_Y, L=l) \nonumber \\ 
  &  \prod_{j=0}^{s}  \big[ \Pr(D_{j+1}=0 \mid D_{j}= Y_{j}=\bar{C}_{j+1}=0,A = a_D, L=l)  \nonumber \\
      &  \times \Pr(Y_{j}=0 \mid D_{j}= Y_{j-1}=\bar{C}_{j}=0,A = a_Y, L=l) \big] \nonumber \Big] \Pr(L=l), \\
\label{eq: identification censoring L}
\end{align}
where $C_k$ is an indicator of being censored at $k$, see Appendix \ref{sec: proof of identifiability} for details. Alternatively, the identification formula can be derived by drawing a SWIG for the scenario of interest, as suggested in Figure \ref{fig: censoring SWIG}. Hereafter we will use $ \nu_{a_Y,a_D,k} $ to denote the g-formula \eqref{eq: identification L}.

\subsection{Alternative representations of the identification formula}
The g-formula \eqref{eq: identification censoring L} can also be expressed as
\begin{align}
   \sum_{s=0}^{k} \mathbb{E}  & [ W_{C,s}(a_Y) W_{D,s}(a_Y,a_D) (1-Y_{s}) (1-D_{s+1}) Y_{s+1} \mid A=a_Y], 
    \label{eq: alternative id formula 1}
\end{align}
where 
\begin{align*}
W_{D,s} (a_Y,a_D) &= \frac{\prod_{j=0}^{s}  \Pr(D_{j+1}=0 \mid \bar{C}_{j+1}=D_{j}= Y_{j}=0, L,  A = a_D) }{ \prod_{j=0}^{s} \Pr(D_{j+1}=0 \mid \bar{C}_{j+1}=D_{j}= Y_{j}=0, L,  A = a_Y) }, \\
W_{C,s} (a_D) &= \frac{I(\bar{C}_{s+1} =0) }{ \prod_{j=0}^{s}  \Pr(\bar{C}_{j+1}=0 \mid \bar{C}_{j}=D_{j}= Y_{j}=0, L,  A = a_D) }, %I(\bar{C}_{j+1}=0)\prod_{j=0}^{k}  \Pr(\bar{C}_{j+1}=0 \mid \bar{C}_{j}=D_{j}= Y_{j}=0,  A = a_D)
\end{align*}
see Appendix \ref{sec: proof of alternative id} for details. Furthermore, another representation of \eqref{eq: identification censoring L} is
\begin{align}
   \sum_{s=0}^{k}  \mathbb{E}   & \{ W_{C,s}(a_D) W_{Y,s}(a_D,a_Y) (1-Y_{s}) (1-D_{s+1}) Y_{s+1} \mid A=a_D \}, %\mid \bar{C}_{k+1} = 0
    \label{eq: alternative id formula 2}
\end{align}
where $W_{C,s} (a_D)$ is defined as in \eqref{eq: alternative id formula 1} and
\begin{align*}
W_{Y,s} (a_D,a_Y) & = \frac{ \Pr(Y_{s+1}=1 \mid \bar{C}_{s+1}=D_{s+1}= Y_{s}=0, L,  A = a_Y) }{\Pr(Y_{s+1}=1 \mid \bar{C}_{s+1}=D_{s+1}= Y_{s}=0, L,  A = a_D) } \\
& \times \frac{\prod_{j=0}^{s-1}  \Pr(Y_{j+1}=0 \mid \bar{C}_{j+1}=D_{j+1}= Y_{j}=0, L,  A = a_Y) }{ \prod_{j=0}^{s-1} \Pr(Y_{j+1}=0 \mid \bar{C}_{j+1}=D_{j+1}= Y_{j}=0, L,  A = a_D) },
\end{align*}
as formally shown in Appendix \ref{sec: proof of alternative id}. Note that in settings without censoring, $W_{C,s} (a) \equiv 1, a = 0,1$.  Representations \eqref{eq: alternative id formula 1} and \eqref{eq: alternative id formula 2} motivate inverse probability (IP) weighted estimators of the separable effects, as described in Section \ref{sec: estimation separable effects}.

\section{Estimation of separable effects}
\label{sec: estimation separable effects}
To estimate the separable effects, we emphasize that \eqref{eq: identification L} and \eqref{eq: identification censoring L} are functionals of (discrete-time) hazard functions and the density of $L$. Indeed, $\Pr(Y_{k+1}=1 \mid D_{k+1}= Y_{k}=\bar{C}_{k+1}=0,A=a, L=l)$ and $\Pr(D_{k+1}=0 \mid D_{k}= Y_{k}=\bar{C}_{k+1}=0,A=a, L=l)$ are often denoted 'cause specific hazard functions' in the statistical literature. Though the term 'cause specific' is confusing because the causal interpretation of these hazard functions is ambiguous \cite{young2018causal}, we can nevertheless estimate these functions using classical statistical models, such as multiplicative or additive hazard models. Provided that these hazard models are correctly specified, along with $\widehat{\Pr}(L=l)$ \cite{young2011comparative}, we can consistently estimate \eqref{eq: identification censoring L} using a parametric g-formula estimator \cite{robins1986new}. However, we can also derive weighted estimators that rely on fewer model assumptions.

\subsection{Inverse probability weighted estimators}
Motivated by the alternative g-formula representation \eqref{eq: alternative id formula 1}, define
\begin{align*}
\hat{W}_{D,k,i} (a_Y,a_D;\hat{\eta}_D) &= \frac{\prod_{j=0}^{k}  \Pr(D_{j+1}=0 \mid \bar{C}_{j+1}=D_{j}= Y_{j}=0, L_{i},  A = a_D; \hat{\eta}_D) }{ \prod_{j=0}^{k} \Pr(D_{j+1}=0 \mid \bar{C}_{j+1}=D_{j}= Y_{j}=0, L_{i},  A = a_Y ; \hat{\eta}_D) }, \\
\hat{W}_{C,k,i} (a_D;\hat{\eta}_{C}) &= \frac{I(\bar{C}_{k+1} =0) }{ \prod_{j=0}^{k}  \Pr(\bar{C}_{j+1}=0 \mid \bar{C}_{j}=D_{j}= Y_{j}=0, L_{i},  A = a_D; \hat{\eta}_{C}) }, 
\end{align*}
where $\Pr(D_{j+1}=0 \mid \bar{C}_{j+1}=D_{j}= Y_{j}=0, L,  A = a_D; \eta_D)$ is a parametric model for the numerator (and denominator) of $W_{D,k}(a_Y,a_D)$ indexed by parameter $ \eta_D$, and $\hat{\eta}_D$ is a consistent estimator of $\eta_D$ (e.g.\ the MLE), and the terms in $\hat{W}_{C,k,i} (a_D;\hat{\eta}_{C})$ are defined similarly, where $\hat{\eta}_{C}$ is a consistent estimator of $\eta_{C}$. 

Let $\eta_1 = (\eta_D, \eta_{C})$, and define the estimator $\hat{\nu}_{1,a_Y,a_D,k} $ of $\nu_{a_Y,a_D,k} $ as the solution to the estimating equation $\sum_{i=1}^{n}U_{1,k,i}(\nu_{a_Y,a_D,k},\hat{\eta}_1)=0$ with respect to $\nu_{a_Y,a_D,k}$ with% $U_i(\nu,\hat{\eta})=\sum_{k=0}^{K}U_{i,k}(\nu,\hat{\eta})$ and 
\begin{align*}
& U_{1,k,i}(\nu_{a_Y,a_D,k},\hat{\eta}_1) \nonumber \\
= & I(A_i=a_Y) \Big[ \sum_{s=0}^{k} \{ \hat{W}_{1,s,i}(a_Y,a_D;\hat{\eta}_1) Y_{s+1,i} (1-Y_{s,i}) (1-D_{s+1,i}) \} - \nu_{a_Y,a_D,k} \Big],  \nonumber \\
\end{align*}
and $\hat{W}_{1,s,i}(a_Y,a_D; \hat{\eta}_1) = \hat{W}_{D,s,i} (a_Y,a_D;\hat{\eta}_{D}) \hat{W}_{C,s,i}  (a_Y;\hat{\eta}_{C}) $. 

Then, $\hat{\nu}_{1,a_Y,a_D, k} $ is a consistent estimator for $\nu_{a_Y,a_D, k}$ if  the models indexed by elements in $\eta_1$ are correctly specified and $\hat{\eta}_1$ is a consistent estimator for $\eta_1$, which follows because \eqref{eq: identification censoring L} and \eqref{eq: alternative id formula 1} are equal. For example, we can use conventional statistical models for binary outcomes, such as pooled logistic regression models, to estimate the weights $W_{D,k}(a_Y,a_D)$ and $W_{C,k}(a_Y)$. 

Analogous to $\hat{\nu}_{1,a_Y,a_D, k} $, we can derive an estimator based on \eqref{eq: alternative id formula 2}. Suppose 
\begin{align*}
\hat{W}_{Y,k,i}  (a_D,a_Y;\hat{\eta}_{Y}) & = \frac{ \Pr(Y_{k+1}=1 \mid \bar{C}_{k+1}=D_{k+1}= Y_{k}=0, L_{i},  A = a_Y; \hat{\eta}_{Y}) }{\Pr(Y_{k+1}=1 \mid \bar{C}_{j+1}=D_{k+1}= Y_{k}=0, L_{i},  A = a_D; \hat{\eta}_{Y}) } \\
& \times \frac{\prod_{j=0}^{k-1}  \Pr(Y_{j+1}=0 \mid \bar{C}_{j+1}=D_{j+1}= Y_{j}=0, L_{i},  A = a_Y; \hat{\eta}_{Y}) }{ \prod_{j=0}^{k-1} \Pr(Y_{j+1}=0 \mid \bar{C}_{j+1}=D_{j+1}= Y_{j}=0, L_{i},  A = a_D; \hat{\eta}_{Y}) }, \\
\end{align*}
where the terms in $\hat{W}_{Y,k,i}  (a_D,a_Y;\hat{\eta}_{Y})$ are statistical models for binary outcomes and $\hat{\eta}_{Y}$ is a consistent estimator for $\eta_Y$. 

Let $\eta_2 = (\eta_Y,\eta_{C})$, and define the estimator $\hat{\nu}_{2,a_Y,a_D,k}$ of $\nu_{a_Y,a_D,k} $ as the solution to the estimating equation $\sum_{i=1}^{n}U_{2,k,i}(\nu_{a_Y,a_D,k},\hat{\eta}_2)=0$ with respect to $\nu_{a_Y,a_D,k}$, where % $U_i(\nu,\hat{\eta})=\sum_{k=0}^{K}U_{i,k}(\nu,\hat{\eta})$ and 
\begin{align*}
& U_{2,k,i}(\nu_{a_Y,a_D,k},\hat{\eta}_2) \nonumber \\
= & I(A_i=a_D) \Big[ \sum_{s=0}^{k} \{ \hat{W}_{2,s,i}(a_Y,a_D;\hat{\eta}_{2}) Y_{s+1,i} (1-Y_{s,i}) (1-D_{s+1,i}) \} -  \nu_{a_Y,a_D,k} \Big],  \nonumber \\
\end{align*}
and $\hat{W}_{2,s,i}(a_Y,a_D;\hat{\eta}_{2} ) = \hat{W}_{C,s,i} (a_D;\hat{\eta}_{C}) \hat{W}_{Y,s,i} (a_D,a_Y;\hat{\eta}_{Y})$. Analogous to the estimator based on \eqref{eq: alternative id formula 1}, provided that the models indexed by elements in $\eta_2$ are correctly specified and $\hat{\eta}_2$ is a consistent estimator for $\eta_2$, then consistency of $\hat{\nu}_{2,a_Y,a_D,k}$ for $\nu_{a_Y,a_D,k} $ follows because \eqref{eq: identification censoring L} and \eqref{eq: alternative id formula 2} are equal.

In the next section, we use this approach to analyze a randomized trial on prostate cancer therapy. In Appendix \ref{sec app: simulations}, we present simulations, suggesting that the estimators perform satisfactorily in finite samples. The simulations also illustrate that the separable effect can be substantially different than the total effect, and that the estimators may be biased if the dismissible component conditions are violated.

\subsection{Example: A randomized trial of prostate cancer}
\label{sec: full example section}

Consider, as described in Section \ref{sec: DES part 1}, a hypothetical drug that has the same direct effect as DES on prostate cancer mortality (same $A_Y$ component), but lacks any effect on mortality due to other causes (opposite $A_D$ component). Then we can define separable direct effects of treatment DES on prostate cancer mortality $Y_k$ in the presence of competing mortality $D_k$ from other causes.
%That is, let $A_{Y}$ denote the effect on testosterone suppression, assuming that this is the mechanism by which DES reduces prostate cancer death. Let $A_{D}$ be the component that has a direct effect on death due to other causes (say, through synthesis of coagulation factors).
% To disentangle the mechanism by which DES reduces prostate cancer mortality, 
%Consider a hypothetical treatment in which the $A_{Y}$ component is included, but the $A_{D}$ component that influences death due to other causes is omitted. Interestingly, a treatment that solely consists of the component $A_{Y}$ is not far from reality, as Luteinising Hormone Releasing Hormone (LHRH) antagonists or orchidectomy (castration) are frequently used to suppress testosterone production in prostate cancer patients today, and these treatments do not contain estrogen. 
We estimated these separable effects using a parametric g-formula estimator and, for simplicity, one of the inverse probability (IP) weighted estimators ($\hat{\nu}_{1,a_Y,a_D,k}$). We used publicly available data from a randomized trial (http://biostat.mc.vanderbilt.edu/DataSets) \cite{byar1980choice} that has been used in several methodological articles on competing events \cite{harrell1996multivariable,kay1986treatment,fine1999analysing,varadhan2010evaluating}. In total, 502 patients were assigned to 4 different treatment arms. We restrict our analysis to the placebo arm (127 patients) and the high-dose DES arm (125 patients). % (5 mg daily) 

To implement the parametric g-formula estimator, we used pooled logistic regression models to estimate the terms in \eqref{eq: identification censoring L}, in which daily activity function, age group, hemoglobin level and previous cardiovascular disease were included as covariates ($L$ in Figure \ref{fig: censoring}), that is, 
\begin{align}
&  \text{logit} [\Pr(Y_{k}=1 \mid D_{k}= Y_{k-1} =\bar{C}_{k}=0,A, L)] = \theta_{0,k} + \theta_{1}A + \theta_{2}Ak + \theta_{3}Ak^2 + \theta'_{4}L  \nonumber  \\
& \label{eq: regression y} \\
& \text{logit} [\Pr(D_{k}=1 \mid D_{k-1}= Y_{k-1}=\bar{C}_{k}=0,A, L)] = \beta_{0,k} + \beta_{1}A + \beta_{2}Ak + \beta_{3}Ak^2 + \beta'_{4}L, \nonumber \\
& \label{eq: regression d}
\end{align}
where $\theta_{0,k}$ and $\beta_{0,k}$ are time-varying intercepts modeled as cubic polynomials. To allow time-varying treatment effects, we included $\theta_{2}, \theta_{3}, \beta_{2}$ and $\beta_{3}$.

To implement the IP weighted estimator $\hat{\nu}_{1,a_Y,a_D,k}$, we only require the model \eqref{eq: regression d} (similarly, we would only require the model \eqref{eq: regression y} to implement $\hat{\nu}_{2,a_Y,a_D,k}$).   

Both the parametric g-formula and IP weighted estimator gave cumulative incidence estimates under the hypothetical drug that were similar, but not identical, to those under DES treatment. Table \ref{tab: 3 years of follow-up} displays estimates of the 3-year cumulative incidence and 95\% bootstrap confidence intervals based on both estimators and Figure \ref{fig: ci plot}B shows cumulative incidence curves from the IP weighted estimator (\texttt{R} code is provided found in the supplementary material).
%the parametric g-formula and the IP weighted estimator $\hat{\nu}_{1,a_Y,a_D,k}$

\begin{table}[ht]
\centering
\caption{Estimates of cumulative incidence after 3 years of follow-up.}
\label{tab: 3 years of follow-up}
\begin{tabular}{|l|r|r|}
\hline
\textbf{Estimand} & \textbf{G-formula estimate (95\%CI)} & $\textbf{IP weighted estimate}$ \textbf{ (95\%CI)} \\ 
  \hline
   $\Pr(Y^{a=1}_{36}=1)$ & 0.14 (0.08-0.20) & 0.17 (0.10,  0.24) \\
  \hline
   $\Pr(Y^{a_Y=1,a_D=0}_{36}=1)$ & 0.15 (0.09-0.21) & 0.18 (0.10, 0.26) \\
   \hline
   $\Pr(Y^{a=0}_{36}=1)$ & 0.21 (0.15-0.28) & 0.23 (0.17,  0.35) \\
   \hline
\end{tabular}
\end{table}

Our analysis suggests that DES mostly reduces prostate cancer mortality via testosterone suppression because the estimate of the separable indirect effect on 3-year mortality is close to zero. Using either the parametric g-formula or the IP weighted estimator, the estimate of the additive indirect effect after 3 years of follow-up is $0.01$ ($0.15-0.14=0.01$ and $0.18-0.17=0.01$), which can be interpreted as the reduction in prostate cancer mortality under DES compared with placebo that is due to the DES effect on mortality from other causes. That is, the total effect of DES on prostate cancer mortality is not simply a consequence of a harmful effect on death from other causes.

The validity of our estimates relies on the assumption that $L$ is sufficient to adjust for the common causes of $Y_k$ and $D_k$. This assumption would be violated if other factors, such as unmeasured comorbidities, exert effects on both $Y_k$ and $D_k$. Also, our approach relies on the absence of time-varying common causes of the event of interest and the competing event in many settings. In future work, we will generalize our approach to allow for time-varying covariates. %become more plausible. 

\section{Discussion}
\label{sec: discussion}
We have defined separable effects as new estimands to promote causal reasoning in competing event settings. The separable effects are motivated by hypothetical interventions, in which a time-fixed treatment is decomposed into distinct components, and each component can be assigned different values.

Therefore, to define and interpret the separable effects, investigators must use their subject-matter knowledge to explicitly articulate a hypothetical decomposition of the treatment. An explicit consideration of this decomposition helps assess the plausibility of the assumptions and guides the design of future experiments to empirically verify the effects \cite{robins2010alternative}.
%\textcolor{red}{Thus, the interpretation of the cumulative incidence functions of the event of interest is difficult, even in an ideal randomized trial in which we also evaluate the cumulative incidence function of the competing event.}

Classical statistical estimands fail to provide the same information as the separable effects (see Young et al \cite{young2018causal} for a detailed discussion of interpretation and identification of counterfactual contrasts in classical estimands for competing event settings). In particular, the cumulative incidence functions of the event of interest and the competing event do not clarify the mechanism by which treatment exerts effects on the event of interest, even if these outcomes are considered jointly in an ideal randomized trial. Furthermore, estimands on the hazard scale, e.g.\ subdistribution hazards and cause-specific hazards, do not have a straightforward causal interpretation and thus cannot solve the problem \cite{young2018causal, hernan2010hazards}.

%interpretation of the cumulative incidence functions of the event of interest is difficult, even in an ideal randomized trial in which we also evaluate the cumulative incidence function of the competing event.

%As an alternative to our separable direct effects, the survivor average causal effect (SACE) can be used in competing risks settings \cite{tchetgen2014identification}. The SACE is equivalent to the total treatment effect, but it is restricted to the principal stratum of patients who would never experience the competing event under either level of treatment \cite{young2018causal,robins1986new, frangakis2002principal}. Unlike the total effect, however, identification of the SACE requires strong untestable assumptions even in a perfectly executed trial. Also, the SACE could never, even in principle, be confirmed in a real-world experiment as it will never be possible to observe the status of the competing event for the same individual under two different levels of treatment. We emhpasise that our separable direct effects are not affected by these shortcomings. 

Identification of separable effects requires, even in a perfectly executed randomized trial, adjustment for pretreatment variables that are common causes of the event of interest and the competing event. However, this strong condition is also needed for the causal interpretation of analysis of trials targeting conventional estimands such as controlled direct effects or counterfactual contrasts of hazard functions \cite{young2018causal}.

For simplicity, we have considered settings in which the treatment $A$ is randomly assigned. For example, we illustrated the application of standard time-to-event methods to estimate the separable effects in a prostate cancer randomized trial. However, our approach can be easily extended to analyses of observational studies under the additional assumption of no unmeasured confounding for the effect of treatment on both the competing event and the event of interest. 

Finally, the idea of separable effects is not only relevant to settings in which the outcome of interest is a time-to-event. Many practical settings involve intermediate outcomes that are ill-defined after the occurrence of a terminating event. For example, we may be interested in treatment effects on outcomes such as quality of life or cognitive function, and these outcomes are meaningless after death. We aim to study separable effects in such settings in future research.

\section*{Acknowledgements}
This work was funded by NIH grant R37 AI102634. M.J.S. was also supported by an ASISA Fellowship and the Research Council of Norway, grant NFR239956/F20.

\bibliography{references}
\bibliographystyle{unsrt}

\clearpage

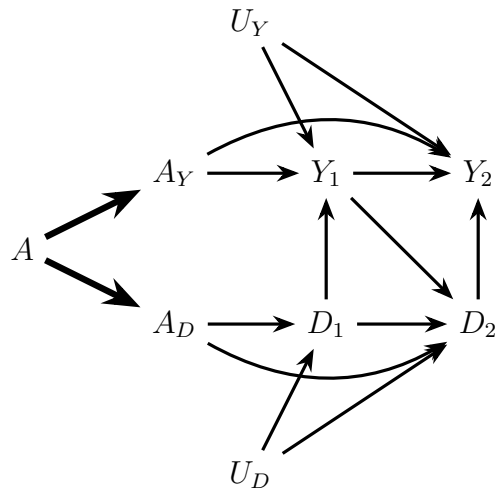
\begin{figure}
\centering
\begin{tikzpicture}
\begin{scope}[every node/.style={thick,draw=none}]
    \node (A) at (0,0) {$A$};
    \node (Ay) at (2,1) {$A_{Y}$};
	\node (Ad) at (2,-1) {$A_{D}$};
	\node (Y1) at (4,1) {$Y_1$};
    \node (D1) at (4,-1) {$D_1$};
    \node (Y2) at (6,1) {$Y_2$};
    \node (D2) at (6,-1) {$D_2$};
    \node (U) at (3,3) {$U_Y$};
   % \node (D3) at (8,-1) {$D(3)$};
    \node (W) at (3,-3) {$U_D$};
\end{scope}

\begin{scope}[>={Stealth[black]},
              every node/.style={fill=white,circle},
              every edge/.style={draw=black,very thick}]
    \path [->] (A) edge[line width=0.85mm] (Ad);
    \path [->] (A) edge[line width=0.85mm] (Ay);
	\path [->] (Ad) edge (D1);
    \path [->] (Ad) edge[bend right] (D2);
	\path [->] (Ay) edge[bend left] (Y2);
    \path [->] (Ay) edge (Y1);	
    \path [->] (W) edge (D1);
    \path [->] (W) edge (D2);
    \path [->] (U) edge (Y1);
    \path [->] (U) edge (Y2);
    \path [->] (Y1) edge (D2);
    \path [->] (Y1) edge (Y2);
    \path [->] (D1) edge (D2);
    \path [->] (D1) edge (Y1);
%    \path [->] (D1) edge (Y2);
    \path [->] (D2) edge (Y2);
\end{scope}
\end{tikzpicture}
\caption{Extension of the directed acyclic graph in Figure 1 which includes unmeasured common causes $U_Y$ and $U_D$, which are expected to exist but do not violate conditions $\Delta$1 and $\Delta$2.}
\label{Figure:DAGfrailty}
\end{figure}

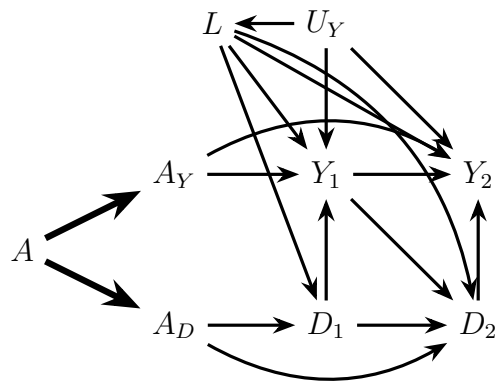
\begin{figure}
\centering
\begin{tikzpicture}
\begin{scope}[every node/.style={thick,draw=none}]
    \node (A) at (0,0) {$A$};
    \node (Ay) at (2,1) {$A_{Y}$};
	\node (Ad) at (2,-1) {$A_{D}$};
	\node (Y1) at (4,1) {$Y_1$};
    \node (D1) at (4,-1) {$D_1$};
    \node (Y2) at (6,1) {$Y_2$};
    \node (D2) at (6,-1) {$D_2$};
    \node (L) at (2.5,3) {$L$};
    \node (U) at (4,3) {$U_{Y}$};
  %  \node (D3) at (8,-1) {$D(3)$};
\end{scope}

\begin{scope}[>={Stealth[black]},
              every node/.style={fill=white,circle},
              every edge/.style={draw=black,very thick}]
    \path [->] (A) edge[line width=0.85mm] (Ad);
    \path [->] (A) edge[line width=0.85mm] (Ay);
	\path [->] (Ad) edge (D1);
    \path [->] (Ad) edge[bend right] (D2);
	\path [->] (Ay) edge[bend left] (Y2);
    \path [->] (Ay) edge (Y1);	
    \path [->] (L) edge (Y1);
    \path [->] (L) edge (Y2);
    \path [->] (L) edge (D1);
    \path [->] (L) edge[bend left] (D2);
    \path [->] (Y1) edge (D2);
    \path [->] (Y1) edge (Y2);
    \path [->] (D1) edge (D2);
    \path [->] (D1) edge (Y1);
    \path [->] (D2) edge (Y2);
    \path [->] (U) edge (Y1);
    \path [->] (U) edge (Y2);
    \path [->] (U) edge (L);
\end{scope}
\end{tikzpicture}
%\caption{Conventional causal diagram of the MR design}
\caption{Conditions $\Delta$1 and $\Delta$2 are valid if the common cause $L$ of $Y_k$ and $D_k$ is measured.}
\label{Figure:DAGconfounder1}
\end{figure}

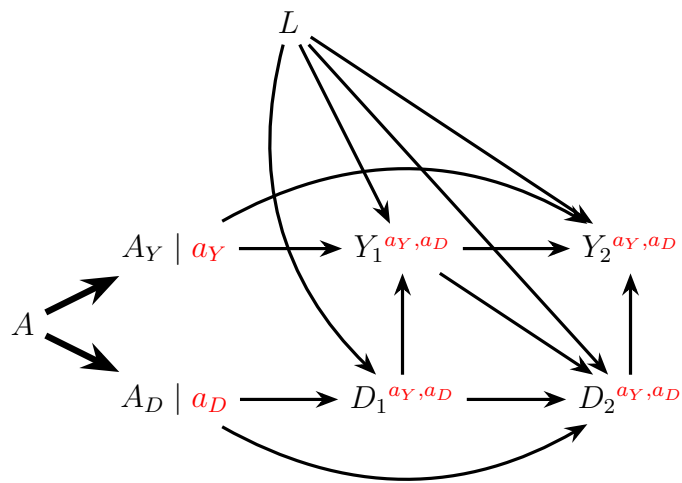
\begin{figure}
\centering
\begin{tikzpicture}
\begin{scope}[every node/.style={thick,draw=none}]
    \node (A) at (-1,0) {$A$};
    \node (Ay) at (1,1) {$A_{Y} \mid \color{red} a_Y$};
	\node (Ad) at (1,-1) {$A_{D} \mid \color{red} a_D$};
	\node (Y1) at (4,1) {$Y_1 \color{red} ^{a_Y,a_D} $};
    \node (D1) at (4,-1) {$D_1 \color{red} ^{a_Y,a_D} $};
    \node (Y2) at (7,1) {$Y_2 \color{red} ^{a_Y,a_D}  $};
    \node (D2) at (7,-1) {$D_2 \color{red} ^{a_Y,a_D}  $};
    \node (L) at (2.5,4) {$L$};
  %  \node (D3) at (8,-1) {$D(3)$};
\end{scope}

\begin{scope}[>={Stealth[black]},
              every node/.style={fill=white,circle},
              every edge/.style={draw=black,very thick}]
    \path [->] (A) edge[line width=0.85mm] (Ad);
    \path [->] (A) edge[line width=0.85mm] (Ay);
	\path [->] (Ad) edge (D1);
    \path [->] (Ad) edge[bend right] (D2);
	\path [->] (Ay) edge[bend left] (Y2);
    \path [->] (Ay) edge (Y1);	
    \path [->] (L) edge (Y1);
    \path [->] (L) edge (Y2);
    \path [->] (L) edge[bend right] (D1);
    \path [->] (L) edge (D2);
    \path [->] (Y1) edge (D2);
%    \path [->] (D1) edge (Y2);
    \path [->] (Y1) edge (Y2);
    \path [->] (D1) edge (D2);
    \path [->] (D1) edge (Y1);
    \path [->] (D2) edge (Y2);
%    \path [->] (D1) edge[bend right] (D3);
%    \path [->] (D2) edge (D3);
%    \path [->] (L) edge[bend left] (D3);
%    \path [->] (Y2) edge (D3);
%    \path [->] (Ad) edge[bend right] (D3);
\end{scope}
\end{tikzpicture}
\caption{Single world intervention template (SWIT) that describes a scenario with interventions on $A_{Y}$ and $A_{D}$.The unmeasured common causes $U_Y$ and $U_D$ have been omitted to avoid clutter.}
\label{fig: simple SWIG}
\end{figure}

%%%%%%%%%%%%%%%%%%%% FIGURE CENSORING
\begin{figure}
\centering
\begin{tikzpicture}
\begin{scope}[every node/.style={thick,draw=none}]
    \node (A) at (0,0) {$A$};
    \node (Ay) at (2,1) {$A_{Y}$};
	\node (Ad) at (2,-1) {$A_{D}$};
	\node (Y1) at (5,1) {$Y_1$};
    \node (D1) at (5,-1) {$D_1$};
    \node (Y2) at (8,1) {$Y_2$};
    \node (D2) at (8,-1) {$D_2$};
    \node (L) at (3,4) {$L$};
    \node (C1) at (4,-3) {$C_1$};
    \node (C2) at (7,-3) {$C_2$};
  %  \node (D3) at (8,-1) {$D(3)$};
\end{scope}

\begin{scope}[>={Stealth[black]},
              every node/.style={fill=white,circle},
              every edge/.style={draw=black,very thick}]
    \path [->] (A) edge[line width=0.85mm] (Ad);
    \path [->] (A) edge[line width=0.85mm] (Ay);
	\path [->] (Ad) edge (D1);
    \path [->] (Ad) edge[bend right] (D2);
	\path [->] (Ay) edge[bend left] (Y2);
    \path [->] (Ay) edge (Y1);	
    \path [->] (L) edge (Y1);
    \path [->] (L) edge (Y2);
    \path [->] (L) edge (D1);
    \path [->] (L) edge (D2);
    \path [->] (Y1) edge (D2);
%    \path [->] (D1) edge (Y2);
    \path [->] (Y1) edge (Y2);
    \path [->] (D1) edge (D2);
    \path [->] (D1) edge (Y1);
    \path [->] (D2) edge (Y2);
    \path [->] (L) edge (C1);
    \path [->] (L) edge[bend right] (C2);
    \path [->] (D1) edge (C2);
    \path [->] (Y1) edge (C2);
    \path [->] (C1) edge (C2);
    \path [->] (Ad) edge (C1);
    \path [->] (Ad) edge (C2);
    \path [->] (Ay) edge (C1);
    \path [->] (Ay) edge[bend right] (C2);
%    \path [->] (D1) edge[bend right] (D3);
%    \path [->] (D2) edge (D3);
%    \path [->] (L) edge[bend left] (D3);
%    \path [->] (Y2) edge (D3);
%    \path [->] (Ad) edge[bend right] (D3);
\end{scope}
\end{tikzpicture}
\caption{Directed acyclic graph with loss to follow-up ($C_k$). We have omitted arrows from $C_k$ into $Y_k$ and from $C_k$ into $D_k$ for $k \in \{1,2\}$.}
\label{fig: censoring}
\end{figure}
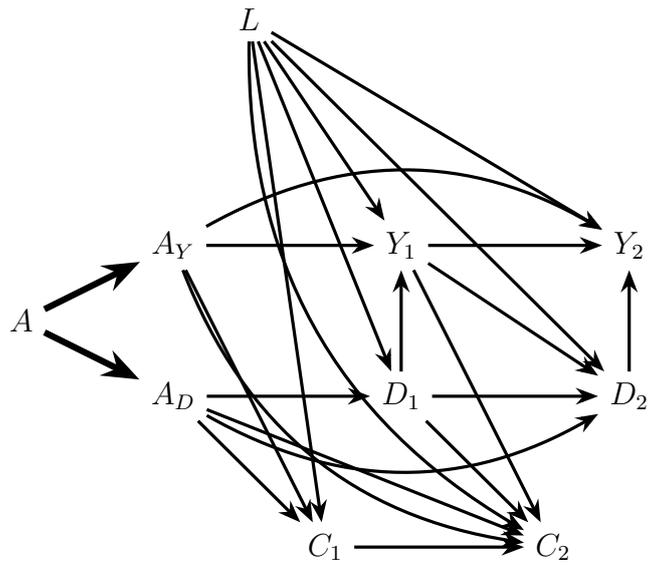

\clearpage
%%%%%%%%%%%%%%%%%%%% FIGURE CENSORING SWIG
\begin{figure}
\centering
\begin{tikzpicture}
\begin{scope}[every node/.style={thick,draw=none}]
    \node (A) at (-1,0) {$A$};
    \node (Ay) at (1,1) {$A_{Y} \mid \color{red} a_Y$};
	\node (Ad) at (1,-1) {$A_{D} \mid \color{red} a_D$};
	\node (Y1) at (4,1) {$Y_1 \color{red} ^{a_Y,a_D,\bar{c}_1}$};
    \node (D1) at (4,-1) {$D_1 \color{red} ^{a_Y,a_D,\bar{c}_1}$};
    \node (Y2) at (7,1) {$Y_2 \color{red} ^{a_Y,a_D,\bar{c}_2}$};
    \node (D2) at (7,-1) {$D_2 \color{red} ^{a_Y,a_D,\bar{c}_2}$};
    \node (L) at (2.5,4) {$L$};
    \node (C1) at (3,-4) {$C_{1}  \mid \color{red} c_1 $};
   \node  (C2) at (6,-4) {$ C_{2} \mid \color{red} c_2 $};
  %  \node (D3) at (8,-1) {$D(3)$};
\end{scope}

\begin{scope}[>={Stealth[black]},
              every node/.style={fill=white,circle},
              every edge/.style={draw=black,very thick}]
    \path [->] (A) edge[line width=0.85mm] (Ad);
    \path [->] (A) edge[line width=0.85mm] (Ay);
	\path [->] (Ad) edge (D1);
    \path [->] (Ad) edge[bend right] (D2);
	\path [->] (Ay) edge[bend left] (Y2);
    \path [->] (Ay) edge (Y1);	
    \path [->] (L) edge (Y1);
    \path [->] (L) edge (Y2);
    \path [->] (L) edge[bend right] (D1);
    \path [->] (L) edge (D2);
    \path [->] (Y1) edge (D2);
%    \path [->] (D1) edge (Y2);
    \path [->] (Y1) edge (Y2);
    \path [->] (D1) edge (D2);
    \path [->] (D1) edge (Y1);
    \path [->] (D2) edge (Y2);
    \path [->] (L) edge (C1);
    \path [->] (L) edge[bend left] (C2);
    \path [->] (D1) edge (C2);
    \path [->] (Y1) edge (C2);
    \path [->] (C1) edge (C2);
    \path [->] (Ad) edge[bend left] (C1);
    \path [->] (Ad) edge (C2);
    \path [->] (Ay) edge[bend left] (C1);
    \path [->] (Ay) edge (C2);
%    \path [->] (D1) edge[bend right] (D3);
%    \path [->] (D2) edge (D3);
%    \path [->] (L) edge[bend left] (D3);
%    \path [->] (Y2) edge (D3);
%    \path [->] (Ad) edge[bend right] (D3);
\end{scope}
\end{tikzpicture}
\caption{Single world intervention graph (SWIG) that describes a scenario with interventions on $A_{Y}$, $A_{D}$ and $\bar{C}_k$. We have omitted arrows from $C_k$ into $Y_k$ and from $C_k$ into $D_k$ for $k \in \{1,2\}$.}
\label{fig: censoring SWIG}
\end{figure}
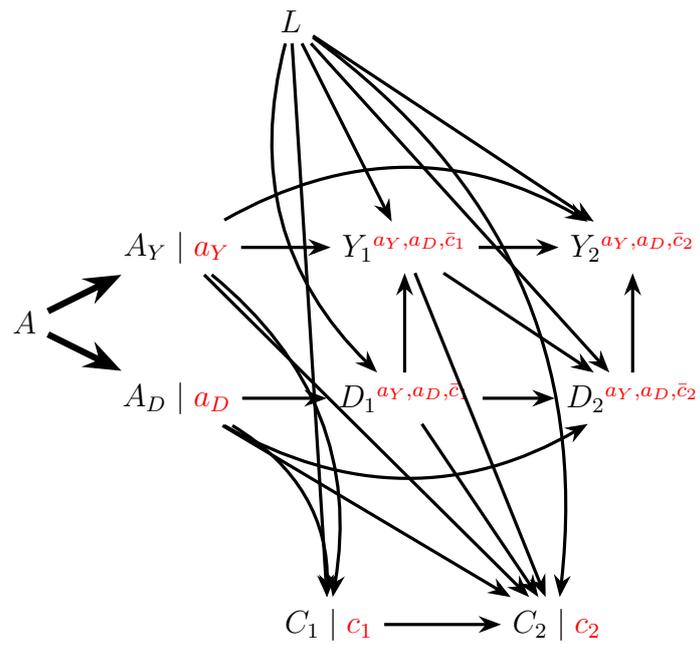

\clearpage
\begin{figure}
\begin{minipage}{.5\linewidth}
\centering
\subfloat[]{\label{main:a}\includegraphics[scale=.55]{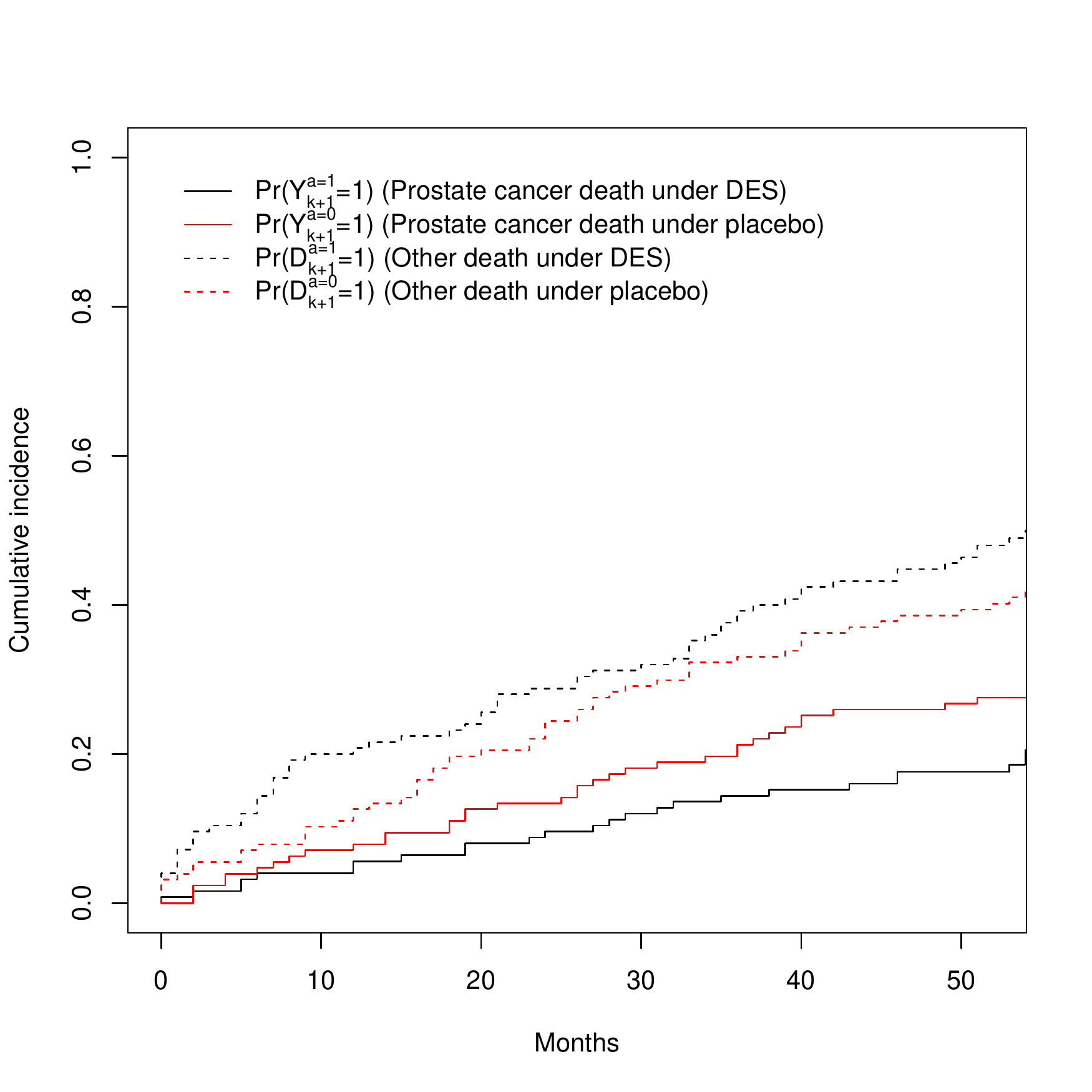}}
\end{minipage}
\begin{minipage}{.5\linewidth}
\centering
\subfloat[]{\label{main:b}\includegraphics[scale=.55]{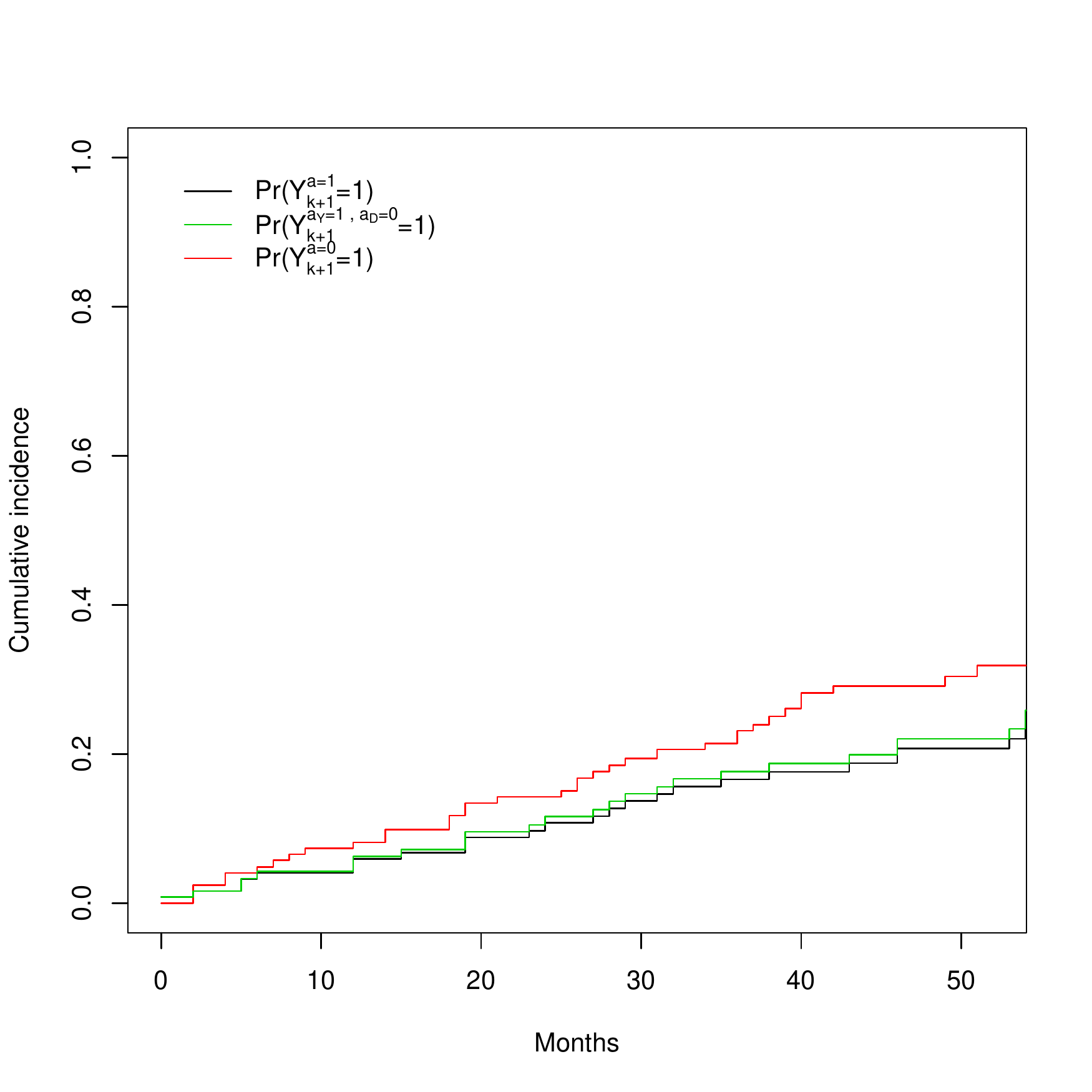}}
\end{minipage}
\caption{Estimated cumulative incidence of (A) death from prostate cancer and death from other causes using the Aalen-Johansen estimator, and (B) death from prostate cancer under DES (black), placebo (red) and the hypothetical treatment where the effect on death of other causes is removed (green) using the inverse probability weighted estimator $\hat{\nu}_{1,a_Y,a_D,k}$.}
\label{fig: ci plot}
\end{figure}

\clearpage
\appendix
\section{Some intuition about the magnitude of the separable direct effects.}
\label{sec app: intuition sep eff}
Consider the following scenarios: 
\begin{itemize}
    \item Scenario 1: $A$ has a null direct effect on the competing event ($ A \nrightarrow D_k$), and the separable direct effect is equal to the total effect.  
\item Scenario 2: $A$ has a null direct effect on the event of interest ($ A \nrightarrow Y_k$), and the indirect effect is equal to the total effect.
\item Scenario 3: $A$ has an average harmful (positive) total effect on both $Y_k$ and $D_k$. The separable direct effects $\Pr (Y_{k+1}^{a_Y=1,a_D}=1)\text{ vs. }\Pr
(Y_{k+1}^{a_Y=0,a_D}=1)$ are harmful (positive), and the separable indirect effects $\Pr (Y_{k+1}^{a_Y,a_D=1}=1)\text{ vs. }\Pr
(Y_{k+1}^{a_Y,a_D=0}=1)$ are protective (negative).
\item Scenario 4: $A$ has an average harmful (positive) total effect on $Y_k$ and a protective (negative) total effect on $D_k$, and the separable direct effects $\Pr (Y_{k+1}^{a_Y=1,a_D}=1)\text{ vs. }\Pr
(Y_{k+1}^{a_Y=0,a_D}=1)$ are harmful (positive), and the separable indirect effects $\Pr (Y_{k+1}^{a_Y,a_D=1}=1)\text{ vs. }\Pr
(Y_{k+1}^{a_Y,a_D=0}=1)$ are harmful (positive).
\end{itemize}

To provide some intuition about the magnitude of the separable effects across these scenarios, we conducted simulations under the following data generating process:
\begin{enumerate}
    \item Draw $L_1 \sim \text{Bernoulli} [p=0.25]$.
    \item Draw $A_Y  \sim \text{Bernoulli} [p=0.5].$
    \item Draw $A_D  \sim \text{Bernoulli} [p=0.5].$
    \item Define $A =a $ if $A_Y=a$ and $A_D = a$.
    \item Set $D_0=Y_0=0$. 
    \item For each $k \in \{0,K\}$, 
    \begin{itemize}
        \item if $D_k=Y_k=0$, \\
    draw $D_{k+1} \sim \text{Bernoulli} [p= \psi_k(A_Y,A_D,L_1,L_2)]$, where
    \begin{align*}
    \psi_k(A_Y,L_1) = \text{expit} & (\omega_{0} + \omega_{1,k}k  + \omega_{2}A_Y +  \omega_{3}L_1)
    %\label{data gen: d model}
    \end{align*} \\
    if $D_{k+1}=0$, \\
    draw $Y_{k+1} \sim \text{Bernoulli} (p= \lambda_k(A_D,L_1))$, where
    \begin{align*}
    \lambda_k(A_D,L_1) = \text{expit} & (\xi_{0} + \xi_{1,k}k  + \xi_{2}A_D + \xi_{3}L_1) % \label{data gen: y model}
    \end{align*} \\
    if $D_{k+1}=1$, set $Y_{k+1}=0$.
    \item else, define $D_{k+1}=D_{k}$,$Y_{k+1}=Y_{k}$. 
    \end{itemize}
\end{enumerate}

\begin{figure}
\begin{minipage}{.45\linewidth}
\centering
\subfloat[]{\label{main:a}\includegraphics[scale=.45]{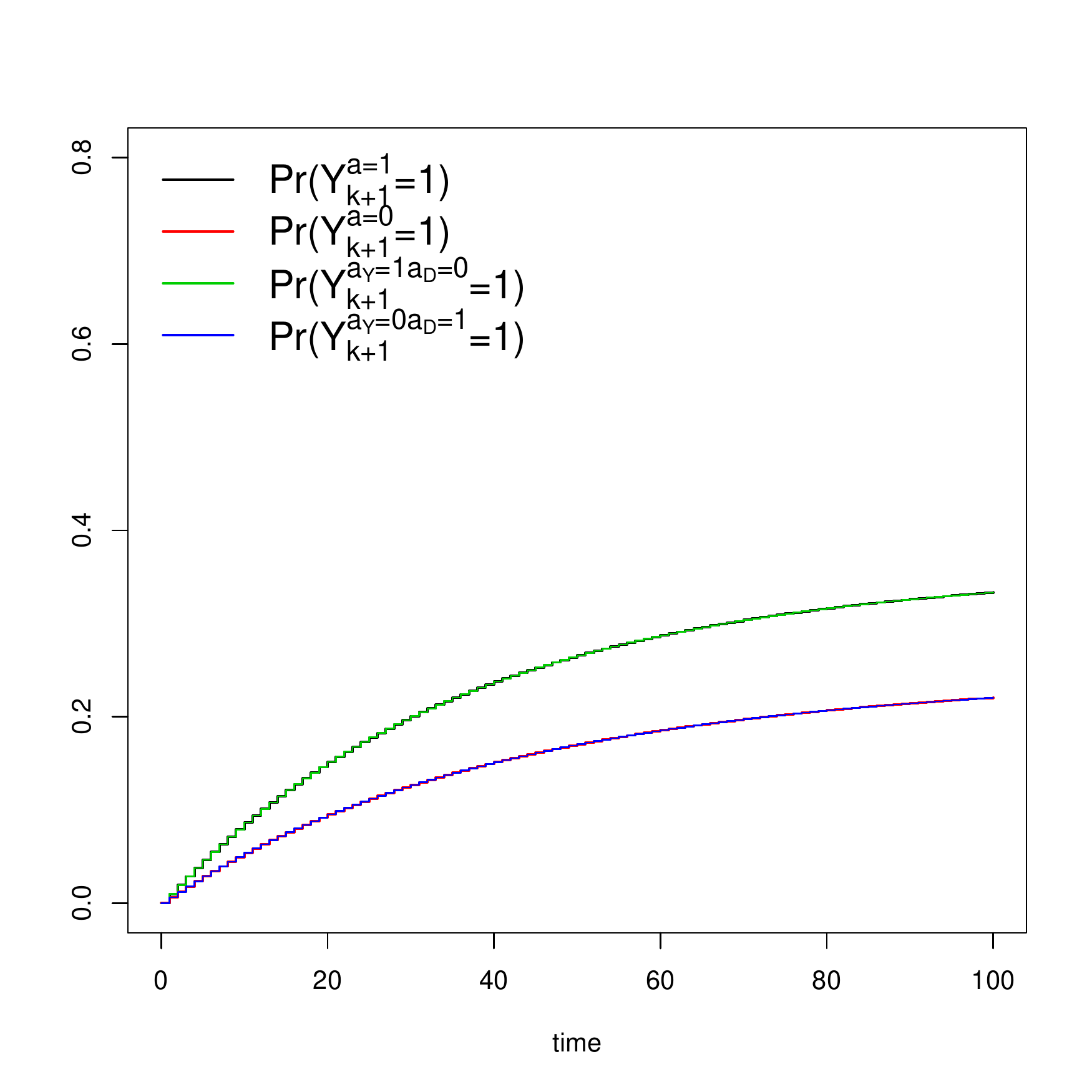}}
\end{minipage}
\begin{minipage}{.45\linewidth}
\centering
\subfloat[]{\label{main:a}\includegraphics[scale=.45]{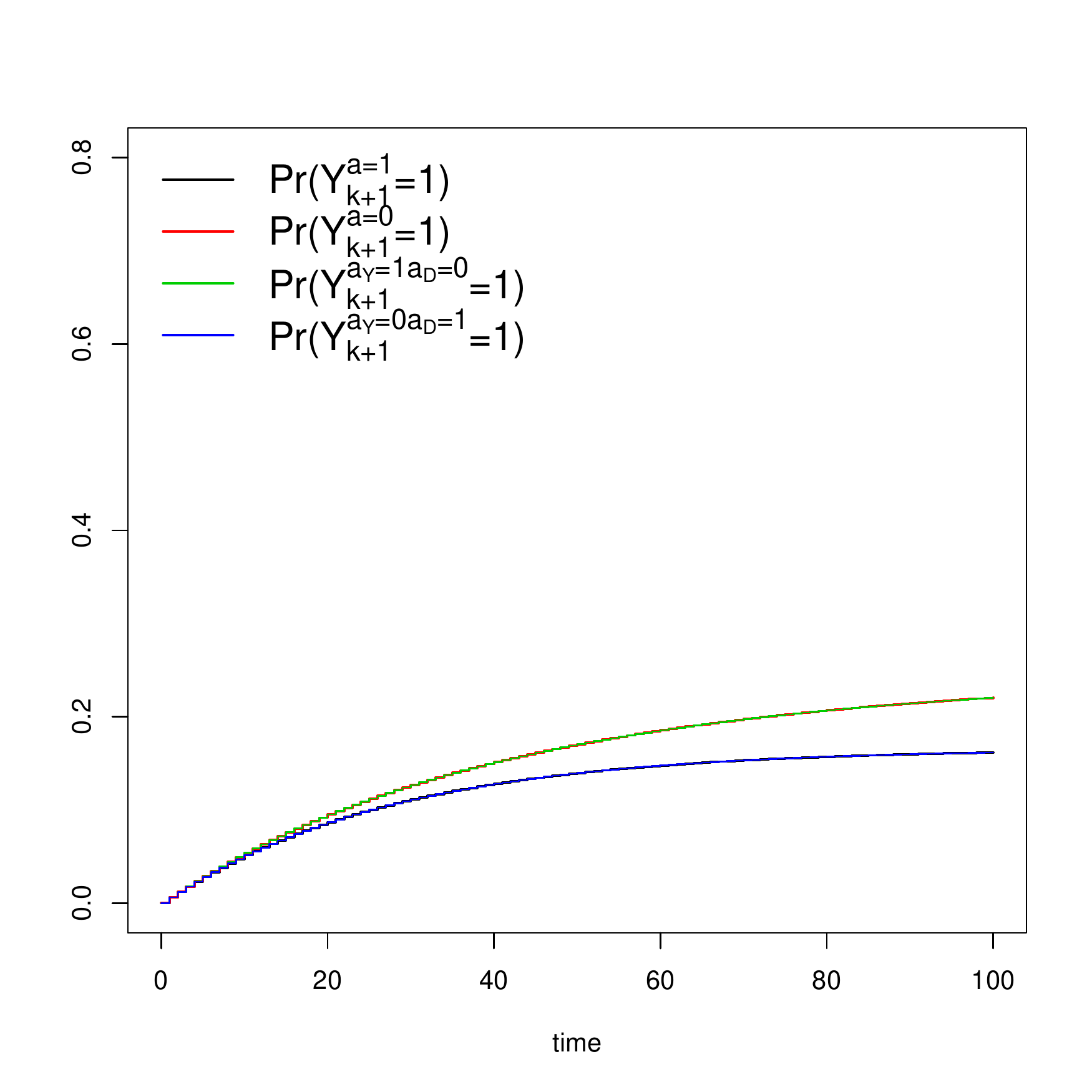}}
\end{minipage}
\begin{minipage}{.45\linewidth}
\centering
\subfloat[]{\label{main:a}\includegraphics[scale=.45]{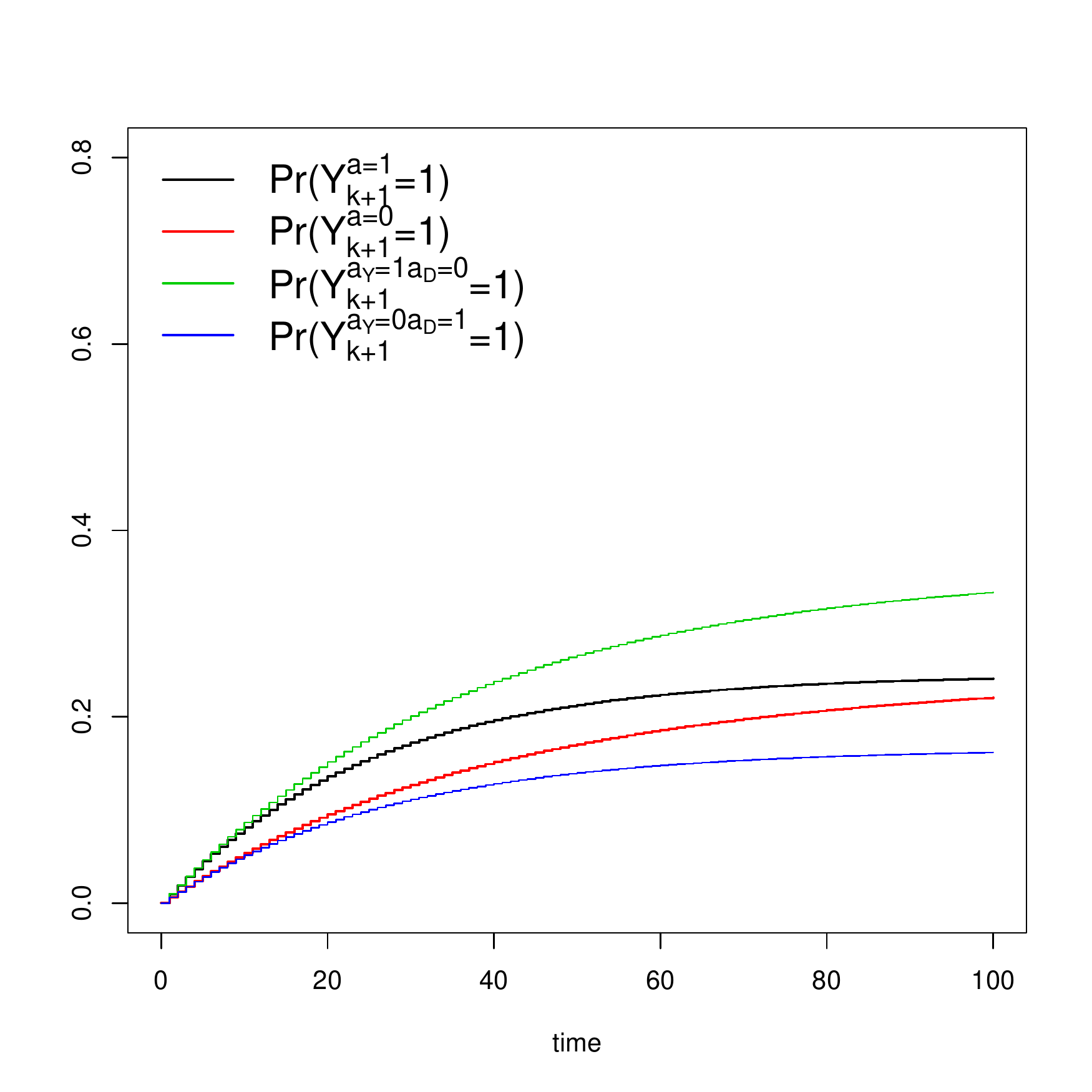}}
\end{minipage}
\begin{minipage}{.45\linewidth}
\centering
\subfloat[]{\label{main:a}\includegraphics[scale=.45]{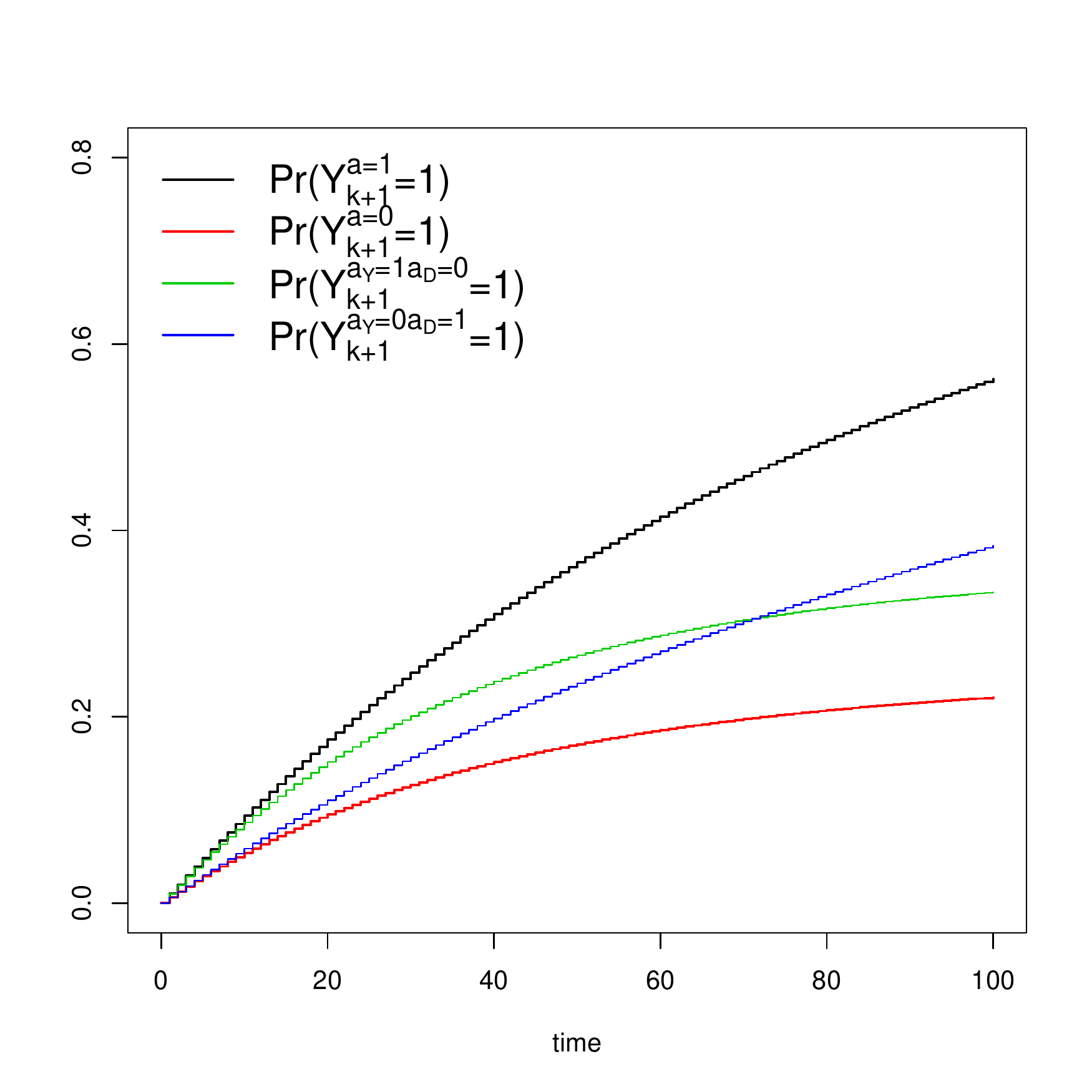}}
\end{minipage}

\caption{Counterfactual outcomes under the data generating mechanisms from Table \ref{tab: sim magnitude illustration}. In the upper left panel, there is perfect overlap between the black and green curves, and of the red and blue curves. In the upper right panel, there is perfect overlap between the red and green curves, and of the black and blue curves.}
%In the upper left panel, there is perfect overlap for $\Pr(Y_{k+1}^{a=1}=1)$ (black) and $\Pr(Y_{k+1}^{a_Y=1,a_D=0}=1)$ (green), as well as $\Pr(Y_{k+1}^{a=0}=1)$ (red) and $\Pr(Y_{k+1}^{a_Y=0,a_D=1}=1)$ (blue). In the upper right panel, there is perfect overlap for $\Pr(Y_{k+1}^{a=0}=1)$ (red) and $\Pr(Y_{k+1}^{a_Y=1,a_D=0}=1)$ (green), as well as $\Pr(Y_{k+1}^{a=1}=1)$ (black) and $\Pr(Y_{k+1}^{a_Y=0,a_D=1}=1)$ (blue).  
\label{fig: illustrative scen}
\end{figure}

Scenario 1 is illustrated in Figure \ref{fig: illustrative scen}a, which was generated using the coefficients from the first row of Table \ref{tab: sim magnitude illustration}.  

Scenario 2 illustrated in Figure \ref{fig: illustrative scen}b, which was generated using the coefficients  from the second row of Table \ref{tab: sim magnitude illustration}.

Scenario 3 is illustrated in Figure \ref{fig: illustrative scen}c, which was generated using the coefficients from the third row of Table \ref{tab: sim magnitude illustration}. 

Scenario 4 is illustrated in Figure \ref{fig: illustrative scen}d, where data were generated from the forth row of Table \ref{tab: sim magnitude illustration}.

%In Figure XXX, we show counterfactual outcomes in 4 settings in which data were generated by the algorithm above, with models \eqref{data gen: d model} and \eqref{data gen: y model} defined as in Table \ref{tab: sim magnitude illustration}. First, the direct effect is equal to the total effect if $A$ exerts no effects on the competing event, as e.g.\ suggested in Figure XXXa. Analogously, if $A$ exerts no direct effect on the event of interest outside of its effect on the competing event, the indirect effect is equal to the total effect, as Suggested in Figure XXXb.

\clearpage

\begin{table}[ht]
\centering
\begin{tabular}{rrrrrrrrrr}
  \hline
Scenario & $\alpha_Y$ & $\omega_1$ & $\omega_2$ & $\omega_3$ & $\alpha_D$ & $\xi_1$ & $\xi_2$ & $\xi_3$ & \\ 
  \hline
 1 & 0.01 & 0 & 10 & 5& 0.03 & 0 & 0 &  5 \\ 
  2 & 0.01 & 0 & 0 & 5& 0.03 & 0 & 5   & 5 \\ 
   3 & 0.01 & 0 & 10 & 5& 0.03 & 0 & 5  & 5 \\ 
    4 & 0.01 & 0 & 10 & 5& 0.03 & 0 & -5   & 5 \\ 
   \hline
\end{tabular}
\caption{Coefficients for the data generating mechanism of the examples in Appendix \ref{sec app: intuition sep eff}.}
\label{tab: sim magnitude illustration}
\end{table}

\clearpage

To provide additional intuition about the magnitude of the separable effects, it may be helpful to consider two hypothetical sets of individuals (Table \ref{tab: hypothetical scenario}). 

First, define the set $Q_k$ of individuals such that $i \in Q_k$ if $i$ would experience the competing event at time $t_{i} < k$  under full treatment (that is, $A_{Y}=1,A_{D}=1$), and would experience the event of interest at a time $s_{i}$, where $t_{i} < s_{i} < k$, under the hypothetical treatment $A_{Y}=1,A_{D}=0$, see Table \ref{tab: hypothetical scenario}. Heuristically, this happens if the hypothetical treatment delays the competing event such that the event of interest is allowed to occur. If $Q_k$ comprises a large fraction of the population, we would expect the total effect and the separable direct effect to be different, because competing events would make it impossible for the event of interest to occur under full treatment, but not under the hypothetical treatment. 

Second, define the set of individuals $R_k$ such that all individuals $j \in R_k$ experience the competing event at time $t_{j} < k$ under full treatment, but would either experience the competing event at $s_{j}$, where $s_{j} < k$, or not experience any event before $k$ under the hypothetical treatment. That is, the subjects in $R_k$ will not experience the event of interest before $k$ under the hypothetical treatment, regardless of the time at which the competing event occurs. If $R_k$ comprises a large fraction of the population, the total effect and the separable direct effect on the event of interest will be close.

\begin{table}[ht]
\centering
\caption{Outcomes at time $k$ in subgroups $Q_k$ and $R_k$.}
\label{tab: hypothetical scenario}
\begin{tabular}{|l|r|r|}
\hline
\textbf{Treatment} & \textbf{Outcomes at $k$ in $Q_k$} & \textbf{Outcomes at $k$ in $R_k$} \\ 
  \hline
   $A_{Y}=1,A_{D}=1$ & $(Y_k=0, D_k=1)$ & $(Y_k=0, D_k=1)$ \\
  \hline
   $A_{Y}=1,A_{D}=0$ & $(Y_k=1, D_k=0)$ & $(Y_k=0, D_k=1)$ or $(Y_k=0, D_k=0)$   \\
   \hline
\end{tabular}
\end{table}

\clearpage

\section{Conditional Independencies that imply the dismissible component conditions.}
\label{sec: DCC as independences}

We expressed the dismissible component conditions $\Delta$1 and $\Delta$2 in terms of equalities of hazard functions. We now show that these equalities are implied by certain counterfactual independencies that can be read directly off of successive single world transformation of a causal DAG. 

\subsubsection*{Hypothetical trial}
Suppose that each component of $A$ is randomly assigned in a hypothetical 4-arm trial $G$. To indicate that the random variables are defined with respect to $G$, let $A_Y(G)$ and $A_D(G)$ be the value of $A_Y$ and $A_D$ observed under $G$, respectively. We assume that $A_Y(G)$ and $A_D(G)$ are randomized independently of each other to values in $\{0,1\}$, that is $A_Y(G) \independent A_D(G)$. Assume no losses to follow-up. Define the independencies
\begin{align}
& Y_{k+1}(G) \independent A_D(G) \mid A_Y(G), Y_{k}(G)=0,D_{k+1}(G)=0, L(G) \label{eq: independence Y L}, \\
& D_{k+1}(G) \independent A_Y(G) \mid A_D(G), D_{k}(G)=0,Y_{k}(G)=0, L(G). \label{eq: independence D L}
\end{align}

\subsection{Conditions that ensure $\Delta$1 and $\Delta$2}
Since $A_Y(G)$ and $A_D(G)$ are randomly assinged, conditional exchangeability is satisfied in the trial $G$, such that
\begin{align*}
& \bar{Y}_{K+1}^{a_Y,a_D}(G),\bar{D}_{K+1}^{a_Y,a_D} (G) \mathpalette{\protect\independenT}{\perp}A_{Y} (G) ,A_{D} (G) \mid L(G),
\end{align*}%
where $a_Y,a_D \in \{0,1\}$. In the special case where $a_Y=a_D$, this conditional exchangeability condition is the same as the conditional exchangeability condition in the main text.

Furthermore, we assume consistency in $G$, that is, if $A_{Y}=a_Y$ and $A_{D}=a_D$ then
\begin{align*}
& Y_{k+1}^{a_Y,a_D}(G)=Y_{k+1}(G) \\
& D_{k+1}^{a_Y,a_D}(G)=D_{k+1}(G),
\end{align*}
where $a_Y,a_D \in \{0,1\}$. This consistency condition is identical to the consistency condition in the main text when $a_Y=a_D$. 

We assume positivity in $G$, that is,  for all $l \in \mathcal{L}$,
\begin{align}
 & \Pr( L(G)=l)>0\implies   \notag \\
& \quad \Pr (A_Y(G)=a_Y, A_D(G)=a_D \mid  L(G)=l)>0, \text{ for } a_Y, a_D \in \{0,1\},
\label{eq: positivity of A G}
\end{align}%
which holds by design in $G$. %\text{ w.p.1}

\begin{figure}
\centering
\begin{tikzpicture}
\begin{scope}[every node/.style={thick,draw=none}]
    \node (Ay) at (2,1) {$A_{Y}$};
	\node (Ad) at (2,-1) {$A_{D}$};
	\node (Y1) at (4,1) {$Y_1$};
    \node (D1) at (4,-1) {$D_1$};
    \node (Y2) at (6,1) {$Y_2$};
    \node (D2) at (6,-1) {$D_2$};
    \node (L) at (3,4) {$L$};
%    \node (C1) at (4,-3) {$C_1$};
%    \node (C2) at (7,-3) {$C_2$};
\end{scope}

\begin{scope}[>={Stealth[black]},
              every node/.style={fill=white,circle},
              every edge/.style={draw=black,very thick}]
	\path [->] (Ad) edge (D1);
    \path [->] (Ad) edge[bend right] (D2);
	\path [->] (Ay) edge[bend left] (Y2);
    \path [->] (Ay) edge (Y1);	
    \path [->] (L) edge (Y1);
    \path [->] (L) edge (Y2);
    \path [->] (L) edge (D1);
    \path [->] (L) edge (D2);
    \path [->] (Y1) edge (D2);
    \path [->] (Y1) edge (Y2);
    \path [->] (D1) edge (D2);
    \path [->] (D1) edge (Y1);
    \path [->] (D2) edge (Y2);
\end{scope}
\end{tikzpicture}
\caption{Directed acyclic graph describing a trial in which $A_Y$ and $A_D$ are randomized. Here,  $\Delta$1 and $\Delta$2 hold.}
\label{fig: trial G}
\end{figure}
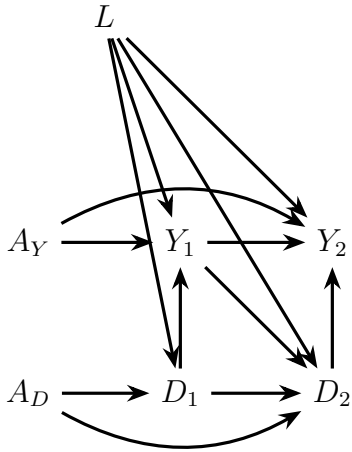

Let $a_Y =0$, $a_D=1$ (an analogous argument holds when  $a_Y =1$, $a_D=0$). Using exchangeability and consistency we find that, for all $l \in \mathcal{L}$,
\begin{align}
    & \Pr (Y_{k+1}(G)=1  \mid Y_{k}(G)=0,D_{k+1}(G)=0, A_Y(G)=0,A_D(G)=1, L(G)=l) \nonumber \\  = & \Pr (Y_{k+1}^{a_Y=0,a_D=1}(G)=1\mid Y_{k}^{a_Y=0,a_D=1}(G)=0,D_{k+1}^{a_Y=0,a_D=1}(G)=0,A_Y(G)=0,A_D=1, L(G)=l) \nonumber \\
    & \qquad \text{ consistency, pos.} \nonumber \\
  = & \Pr (Y_{k+1}^{a_Y=0,a_D=1}(G)=1\mid Y_{k}^{a_Y=0,a_D=1}(G)=0,D_{k+1}^{a_Y=0,a_D=1}(G)=0,L(G)=l) \qquad \text{exchangeability} \nonumber \\
    \label{eq: identical ass 1}
\end{align} Similarly, using \eqref{eq: independence Y L}, exchangeability and consistency we find
\begin{align}
    & \Pr (Y_{k+1}(G)=1  \mid Y_{k}(G)=0,D_{k+1}(G)=0, A_Y(G)=0,A_D(G)=1, L(G)=l) \nonumber \\ 
  =& \Pr (Y_{k+1}(G)=1  \mid Y_{k}(G)=0,D_{k+1}(G)=0, A_Y(G)=0, L(G)=l) \qquad \text{due to \eqref{eq: independence Y L}} \nonumber \\
 =& \Pr (Y_{k+1}(G)=1  \mid Y_{k}(G)=0,D_{k+1}(G)=0, A_Y(G)=0,A_D(G)=0, L(G)=l) \qquad \text{due to \eqref{eq: independence Y L}} \nonumber \\
  = & \Pr (Y_{k+1}^{a_Y=0,a_D=0}(G)=1\mid Y_{k}^{a_Y=0,a_D=0}(G)=0,D_{k+1}^{a_Y=0,a_D=0}(G)=0,A_Y(G)=A_D(G)=0, L(G)=l) \nonumber \\
 & \qquad \text{ consistency, pos.} \nonumber \\
  = & \Pr (Y_{k+1}^{a_Y=0,a_D=0}(G)=1\mid Y_{k}^{a_Y=0,a_D=0}(G)=0,D_{k+1}^{a_Y=0,a_D=0}(G)=0,L(G)=l) \qquad \text{exchangeability} \nonumber \\
\label{eq: identical ass 2}
\end{align} 
The derivations in \eqref{eq: identical ass 1} and \eqref{eq: identical ass 2} show that $\Delta$1 is satisfied if condition \eqref{eq: independence Y L} holds, assuming conditional exchangeability, positivity and consistency. We can use exactly the same argument to show that condition $\Delta$2 holds under conditional exchangeability, positivity, consistency and condition \eqref{eq: independence D L}. Conditions \eqref{eq: independence Y L} and \eqref{eq: independence D L} are helpful in practice because these independences can be evaluated in causal graphs. In particular, these conditions hold in Figure \ref{fig: trial G}, where we have described a trial in which $A_Y$ and $A_D$ are randomly assigned such that $\Pr(A_Y=a_Y,A_D=a_D) > 0$ for all $a_D,a_Y \in \{0,1\}$. 

Note that conditions \eqref{eq: implication 1} and \eqref{eq: implication 2} in the main text, which are part of the decomposition assumption, are required for the independencies \eqref{eq: independence Y L} and \eqref{eq: independence D L} to hold. 

\section{Proof of identifiability}
\label{sec: proof of identifiability}
We assume a Finest Fully Randomized Causally Interpretable Structured Tree Graph (FFRCISTG)  model \cite{robins1986new}. The aim is to identify $P \left( Y^{ a_Y,a_D, \bar{c}=0}_k = 1 \right)$ as a function of the factual data, in which $A$ is randomized. To do this, we will initially consider a scenario in which both $A_{Y}$ and $A_{D}$ are randomized, that is, we consider a 4 arm trial $G$, as described in Appendix \ref{sec: DCC as independences}. Hereafter we omit the string '$(G)$' after the random variables, e.g.\ $A_Y(G) = A_Y$, to avoid clutter. We will provide a proof for the scenario with a measured pretreatment covariate $L$ and censoring $C_k$. The results will immediately hold in simpler scenarios, e.g.\ by defining $L$ or $C_k$ to be the empty set.

\subsection{Identifiabilty conditions in the presence of censoring}
\label{sec: appendix censoring}
First, we generalize the identifiability conditions to allow for censoring. Assume that subjects may be lost to follow-up, and that the losses to follow-up can depend on $A_{Y}$, $A_{D}$ and $L$, as suggested in Figure \ref{fig: censoring}. Further, assume that the losses to follow-up are independent of future counterfactual events ('independent censoring'). To be more precise, we consider a setting in which we intervened such that no subject was lost to
follow-up. Let $C_k \in \{0,1\}$ be an indicator of loss to follow-up by $k$. Let $D^{a_Y,a_D, \bar{c}=0}_k$ and $Y^{a_Y,a_D, \bar{c}=0}_k$ be the counterfactual
values of $Y_k$ and $D_k$ when $A_{Y}$ is set to $a^*$, $A_{D}$ is set to $a$, and
follow-up is ensured at all times.

In a continuous time setting, it is usually assumed that two events cannot occur at the same point in time. In our discrete time setting with pretreatment covariates $L$ and censoring $C_k$, we define a temporal order 
\begin{equation*}
(L,A_D,A_Y,C_{1},D_{1},Y_{1},C_{2},D_{2},Y_{2},...,C_{K+1},D_{K+1},Y_{K+1}).
\end{equation*}
For all $k \in \{0, K\}$ we consider the following conditions. First, we extend the exchangeability conditions from Section \ref{sec: simple id conditions},
\begin{align*}
& \mathbf{E1}: \bar{Y}_{K+1}^{a,\bar{c}=0},\bar{D}_{K+1}^{a,\bar{c}=0} \mathpalette{\protect\independenT}{\perp}
A \mid L  \notag \\
& \mathbf{E2}: \underbar{Y}^{a, \bar{c}=0}_{k+1}, \underbar{D}%
^{a, \bar{c}=0}_{k+1} \mathpalette{\protect\independenT}{\perp}
C_{k+1} \mid Y_k = D_k = \bar{C}_k = 0, L, A.
\end{align*}
Here, as in Section \ref{sec: simple id conditions}, E1 holds when $A \equiv A_{Y} \equiv A_{D}$ are randomized. E2 requires that losses to follow-up are independent of future counterfactual events, given the measured past. This condition is similar to the 'independent censoring' condition that is assumed to hold in classical randomized trials \cite{young2018causal}. \newline

Furthermore, we require a consistency condition such that if $A_{Y}=a_Y$, $A_{D}=a_D$ and $\bar{C}_k = 0$, then $Y_k = {Y}^{a_Y,a_D, \bar{c}=0}_k$ and $D_k = {D}^{a_Y,a_D, \bar{c}=0}_k  $, and still we only observe scenarios where $a_Y=a_D$. The consistency condition ensures that if an individual has a data history consistent with the intervention under a counterfactual scenario, then the
observed outcome is equal to the counterfactual outcome.

Similar to Section \ref{sec: simple id conditions}, the exchangeability and consistency conditions are conventional in the causal inference literature. We also require an extra positivity condition in the presence of censoring, that is, 
\begin{align*}
& \Pr(A=a,Y_k=0,D_k=0,\bar{C}_k=0,L=l) > 0 \implies  \notag \\
&\Pr(C_{k+1} = 0\mid Y_k=0,D_k=0,\bar{C}_k=0,L=l,A=a) > 0, %\text{ w.p.1}
\end{align*}
for $a=\{0,1\}$, which ensures that for any possible history of treatment assignments and covariates among those who are event-free and uncensored at $k$, some subjects will remain uncensored at $k+1$.

Finally, we rely on two dismissible component conditions which generalize the conditions in Section \ref{sec: identification main}, by allowing for a hypothetical intervention to eliminate  censoring at all times. 

\textbf{Dismissible component conditions:} 
For all $l \in \mathcal{L}$,
\begin{align*}
\mathbf{\Delta1_{c}: } & \Pr(Y^{a_Y,a_D=1,\bar{c}= 0}_{k+1} = 1 \mid Y^{a_Y,a_D=1,\bar{c}=
0}_t = 0, D^{a_Y,a_D=1,\bar{c}= 0}_{k+1} = 0, L=l )  \notag \\
&=\Pr(Y^{a_Y,a_D=0,\bar{c}= 0}_{k+1} = 1 \mid Y^{a_Y,a_D=0,c =
0}_t = 0, D^{a_Y,a_D=0,c = 0}_{k+1} = 0 , L=l )
\\
\\
\mathbf{\Delta2_{c}: } & \Pr(D^{a_Y=1,a_D,\bar{c}= 0}_{k+1} = 1 \mid Y^{a_Y=1,a_D,\bar{c}=
0}_k = 0, D^{a_Y=1,a_D,\bar{c}= 0}_k = 0, L=l )  \notag \\
&=\Pr(D^{a_Y=0,a_D,\bar{c}= 0}_{k+1} = 1 \mid Y^{a_Y=0,a_D,c = 0}_k = 0,
D^{a_Y=0,a_D,c = 0}_k = 0 , L=l ).
\end{align*}

Under these conditions, $\Pr(Y^{a_Y,a_D, \bar{c}=0}_{K+1}=1)$ is identified from \eqref{eq: identification censoring L}.

\subsection{Proof of identifiability}
We consider the counterfactual outcomes in a setting where $a_Y=0$ and $a_D=1$ (analogous arguments holds for the setting where $a_Y=1$ and $a_D=0$), and we use laws of probability as well as $\Delta$1$_{c}$ and $\Delta$2$_{c}$ to express 
\begin{align}
  & \Pr(Y^{a_Y=0,a_D=1, \bar{c}=0}_{K+1}=1)  \nonumber \\
  =& \sum_{l}  \Big[  \Pr(Y^{a_Y=0,a_D=1, \bar{c}=0}_{K+1}=1 \mid  L=l) \Big] \Pr(L=l)  \nonumber \\
    =& \sum_{l}  \Big[ \sum_{s=0}^{K} \Pr(Y^{a_Y=0,a_D=1, \bar{c}=0}_{s+1}=1 \mid D^{a_Y=0,a_D=1, \bar{c}=0}_{s+1}= Y^{a_Y=0,a_D=1, \bar{c}=0}_{s}=0, L=l) \nonumber \\ 
   &  \prod_{j=0}^{s}  \big[ \Pr(D^{a_Y=0,a_D=1, \bar{c}=0}_{j+1}=0 \mid D^{a_Y=0,a_D=1, \bar{c}=0}_{j}= Y^{a_Y=0,a_D=1, \bar{c}=0}_{j}=0, L=l)  \nonumber \\
      &  \times \Pr(Y^{a_Y=0,a_D=1, \bar{c}=0}_{j}=0 \mid D^{a_Y=0,a_D=1, \bar{c}=0}_{j}= Y^{a_Y=0,a_D=1, \bar{c}=0}_{j-1}=0, L=l) \big] \Big] \Pr(L=l)  \nonumber \\
       =& \sum_{l}  \Big[ \sum_{s=0}^{K} \Pr(Y^{a_Y=0,a_D=0, \bar{c}=0}_{s+1}=1 \mid D^{a_Y=0,a_D=0, \bar{c}=0}_{s+1}= Y^{a_Y=0,a_D=0, \bar{c}=0}_{s}=0, L=l) \nonumber \\ 
   &  \prod_{j=0}^{s}  \big[ \Pr(D^{a_Y=1,a_D=1 \bar{c}=0}_{j+1}=0 \mid D^{a_Y=1,a_D=1 \bar{c}=0}_{j}= Y^{a_Y=1,a_D=1 \bar{c}=0}_{j}=0, L=l)  \nonumber \\
      &  \times \Pr(Y^{ a_Y=0,a_D=0, \bar{c}=0}_{j}=0 \mid D^{ a_Y=0,a_D=0, \bar{c}=0}_{j}= Y^{ a_Y=0,a_D=0, \bar{c}=0}_{j-1}=0, L=l) \nonumber \big]  \Big] \Pr(L=l), \\
    =& \sum_{l}  \Big[ \sum_{s=0}^{K} \Pr(Y^{a=0, \bar{c}=0}_{s+1}=1 \mid D^{a=0, \bar{c}=0}_{s+1}= Y^{a=0, \bar{c}=0}_{s}=0, L=l) \nonumber \\ 
   &  \prod_{j=0}^{s}  \big[ \Pr(D^{a=1 \bar{c}=0}_{j+1}=0 \mid D^{a=1 \bar{c}=0}_{j}= Y^{a=1 \bar{c}=0}_{j}=0, L=l)  \nonumber \\
      &  \times \Pr(Y^{ a=0, \bar{c}=0}_{j}=0 \mid D^{ a=0, \bar{c}=0}_{j}= Y^{ a=0, \bar{c}=0}_{j-1}=0, L=l) \nonumber \big]  \Big] \Pr(L=l), \\
      \label{eq: step 2 counterfactual relation}
\end{align}
where $Y^{a_Y,a_D, \bar{c}=0}_{-1}$ and $Y^{a_Y, \bar{c}=0}_{-1}$ are empty sets.

For $s \geq 0$ and all $l$ such that $ \Pr(D^{a, \bar{c}=0}_{s+1}= Y^{a, \bar{c}=0}_{s}=0,L=l) > 0 $, let us consider the term
\begin{align*}
 & \Pr(Y^{a, \bar{c}=0}_{s+1}=1 \mid D^{a, \bar{c}=0}_{s+1}= Y^{a, \bar{c}=0}_{s}=0,L=l) \nonumber \\
= & \Pr(Y^{a, \bar{c}=0}_{s+1}=1 \mid D^{a, \bar{c}=0}_{s+1}= Y^{a, \bar{c}=0}_{s}=Y_{0}=D_{0}=\bar{C}_{0}=0,L=l) \nonumber \\
 =& \frac{\Pr(Y^{a, \bar{c}=0}_{s+1}=1, \bar{D}^{a, \bar{c}=0}_{s+1}= \bar{Y}^{a, \bar{c}=0}_{s}=0 \mid Y_{0}=D_{0}=\bar{C}_{0}=0, A=a,L=l)}{P(\bar{D}^{a, \bar{c}=0}_{s+1}= \bar{Y}^{a, \bar{c}=0}_{s}=0 \mid Y_{0}=D_{0}=\bar{C}_{0}=0, A=a,L=l)},  \nonumber \\
\end{align*}
where we use the fact that all subjects are event-free and uncensored at $k=0$ in the 2nd line, and laws of probability and E1 in the 3rd line. Then, we use positivity and E2 to find
\begin{align}
 & \Pr(Y^{a, \bar{c}=0}_{s+1}=1 \mid D^{a, \bar{c}=0}_{s+1}= Y^{a, \bar{c}=0}_{s}=Y_{0}=D_{0}=\bar{C}_{0}=0, A=a,L=l) \nonumber \\
 =& \frac{\Pr(Y^{a, \bar{c}=0}_{s+1}=1, \bar{D}^{a, \bar{c}=0}_{s+1}= \bar{Y}^{a, \bar{c}=0}_{s}=0 \mid Y_{0}=D_{0}=\bar{C}_{1}=0, A=a,L=l)}{P(\bar{D}^{a, \bar{c}=0}_{s+1}= \bar{Y}^{a, \bar{c}=0}_{s}=0 \mid Y_{0}=D_{0}=\bar{C}_{1}=0, A=a,L=l)}  \nonumber \\
 =&\Pr(Y^{a, \bar{c}=0}_{s+1}=1 \mid D^{a, \bar{c}=0}_{s+1}= Y^{a, \bar{c}=0}_{s}=Y_{0}=D_{0}=\bar{C}_{1}=0, A=a,L=l). \nonumber \\
 \label{eq: step 1 pos exch}
\end{align}
Similarly, if $s=1$ we use consistency, a new step like \eqref{eq: step 1 pos exch}, and consistency to find that 
\begin{align*}
 &  \Pr(Y^{a, \bar{c}=0}_{2}=1 \mid D^{a, \bar{c}=0}_{2}= Y^{a, \bar{c}=0}_{1}=Y_{0}=D_{0}=\bar{C}_{1}=0, A=a,L=l) \nonumber \\
 = &  \Pr(Y^{a, \bar{c}=0}_{2}=1 \mid D^{a, \bar{c}=0}_{2}=Y_{1}=D_{1}=\bar{C}_{1}=0, A=a,L=l) \nonumber \\
  = &  \Pr(Y^{a, \bar{c}=0}_{2}=1 \mid D^{a, \bar{c}=0}_{2}=Y_{1}=D_{1}=\bar{C}_{2}=0, A=a,L=l) \nonumber \\
 =&  \Pr(Y_{2}=1 \mid Y_{1}=D_{2}=\bar{C}_{2}=0,  A=a,L=l). \nonumber \\
\end{align*}
If $s> 1$, we use consistency to find 
\begin{align}
 &  \Pr(Y^{a, \bar{c}=0}_{s+1}=1 \mid D^{a, \bar{c}=0}_{s+1}= Y^{a, \bar{c}=0}_{s}=Y_{0}=D_{0}=\bar{C}_{1}=0, A=a,L=l) \nonumber \\
 =&  \Pr(Y^{a, \bar{c}=0}_{s+1}=1 \mid D^{a, \bar{c}=0}_{s+1}= Y^{a, \bar{c}=0}_{s}=Y_{1}=D_{1}=\bar{C}_{1}=0, A=a,L=l). \nonumber \\
 \label{eq: step 2 consistency}
\end{align}

Then, we repeat the steps in \eqref{eq: step 1 pos exch} and \eqref{eq: step 2 consistency} to find that for all $s \in (1,2,...,K+1)$,
\begin{align}
 & \Pr(Y^{a, \bar{c}=0}_{s+1}=1 \mid D^{a \bar{c}=0}_{s+1}= Y^{a, \bar{c}=0}_{s}=Y_{0}=D_{0}=\bar{C}_{0}=0, A=a,L=l) \nonumber \\
      =&\Pr(Y_{s+1}=1 \mid D_{s+1}= Y_{s}=\bar{C}_{s+1}=0, A=a,L=l). \nonumber \\
\label{eq: hazard Y}
\end{align}

Similarly, for $D^{a, \bar{c}=0}_{s+1}$ we could follow the same steps as for $Y^{a, \bar{c}=0}_{s+1}$ to express
\begin{align}
 & \Pr(D^{a, \bar{c}=0}_{s+1}=1 \mid D^{a, \bar{c}=0}_{s}= Y^{a, \bar{c}=0}_{s}=Y_{k}=D_{k}=\bar{C}_{k}=0,A = a, L=l) \nonumber \\
      =&\Pr(D_{s+1}=1 \mid D_{s}= Y_{s}=\bar{C}_{s+1}=0,A = a, L=l). \nonumber \\
\label{eq: hazard D}
\end{align}
Using the results in \eqref{eq: step 2 counterfactual relation}, \eqref{eq: hazard Y} and \eqref{eq: hazard D}, we find that
\begin{align*}
& \Pr(Y^{a_Y,a_D, \bar{c}=0}_{K+1}=1)  \nonumber \\
 =& \sum_{l}  \Big[ \sum_{s=0}^{K} \Pr(Y_{s+1}=1 \mid D_{s+1}= Y_{s}=\bar{C}_{s+1}=0,A =a_Y, L=L=l) \nonumber \\ 
   &  \prod_{j=0}^{s}  \big[ \Pr(D_{j+1}=0 \mid D_{j}= Y_{j}=\bar{C}_{j+1}=0,A = a_D, L=L=l)  \nonumber \\
      &  \times \Pr(Y_{j}=0 \mid D_{j}= Y_{j-1}=\bar{C}_{j}=0,A = a_Y, L=L=l) \big] \nonumber \Big] \Pr(L=L=l). \\
\end{align*}
In words, we have derived that $\Pr(Y^{a_Y,  a_D, \bar{c}=0}_{K+1}=1)$ is identified from a trial in which only subjects with $(A_{Y}= A_{D}= A)$ are observed, i.e.\ in a trial in which $A$ is randomized. Hence, in practice we only need data from the treatment arms in which $A \equiv A_{Y} \equiv A_{D} \in \{0,1\}$. 

\clearpage
\section{Proof of weighted representations}
\label{sec: proof of alternative id}

For the ease of exposition, define
\begin{align*}
    W'_{C,k} (a_Y) = \frac{1 }{ \prod_{j=0}^{k}  \Pr(C_{j+1}=0 \mid \bar{C}_{j}=D_{j}= Y_{j}=0, L = l,  A = a_D) }.
\end{align*}

Consider the expression
\begin{align*}
    \mathbb{E}  & [ W_{C,k}(a_Y) W_{D,k}(a_Y,a_D)  Y_{k+1} (1-Y_{k}) (1-D_{k+1}) \mid A=a_Y] \\ %\mid \bar{C}_{k+1} = 0 \\ 
    =  & \mathbb{E}  [ W'_{C,k} (a_Y) W_{D,k}(a_Y,a_D)  Y_{k+1} (1-Y_{k}) (1-D_{k+1}) (1-\bar{C}_{k+1}) \mid A=a_Y] \\ %\mid \bar{C}_{k+1} = 0 \\ 
    =& \sum_{l}  \sum_{\bar{y}_{k+1}} \sum_{\bar{d}_{k+1}} [ \Pr(\bar{y}_{k+1}, d_{k+1},c_{k+1},l \mid A = a_Y) W'_{C,k} (a) W_{D,k}(a_Y,a_D)  \\
    & \times y_{k+1} (1-y_{k}) (1-d_{k+1}) (1-c_{k+1}) ]  \\ %\textbf{} \text{(by definition of the expected value)}
    =& \sum_{l}  [\Pr(Y_{k+1}=1,Y_k=D_{k+1}=\bar{C}_{k+1}=0,l \mid A = a_Y) W'_{C,k} (a_Y) W_{D,k}(a_Y,a_D) ]  \\
    =& \sum_{l}  [ \Pr(Y_{k+1}=1 \mid Y_k= D_{k+1}=\bar{C}_{k+1}=0,L=l, A=a_Y) \\
    & \times \Pr( D_{k+1}=0 \mid \bar{C}_{k+1}= \bar{D}_k= \bar{Y}_k =0,L=l,A=a_Y )    \\
    & \times \Pr(C_{k+1}= 0 \mid \bar{D}_k= \bar{Y}_{k}= \bar{C}_{k} =0,L=l, A=a_Y)   \\
    & \times \Pr(\bar{Y}_{k}=\bar{D}_{k}=\bar{C}_{k}=0, L=l \mid  a_Y  ) \\
    & \times W'_{C,k} (a_Y) W_{D,k}(a_Y,a_D) ]  \\ 
\end{align*}
where we use the definition of expected value, the fact that $Y_k$ and $D_k$ are binary, and laws of probability.

We use laws of probability to express $ \Pr(\bar{Y}_{k}=\bar{D}_{k}=\bar{C}_{k}=0, l \mid A= a_Y )$ as
\begin{align*}
 &   \Pr(Y_{k}=0 \mid \bar{C}_{k}=D_{k}= Y_{k-1}=0,L=l,  A = a_Y) \nonumber \\
   & \times \Pr(D_{k}=0 \mid \bar{C}_{k}=D_{k-1}= Y_{k-1}=0,L=l,  A = a_Y)  \nonumber \\
    & \times \Pr(C_{k}= 0 \mid D_{k-1} = Y_{k-1}= \bar{C}_{k-1} =0,L=l, A = a_Y)  \\
    & \times \Pr(\bar{Y}_{k-1}=\bar{D}_{k-1}=0, \bar{C}_{k-1}=0, l \mid  A= a_Y ), \\
\end{align*}
where any variable indexed with a number $m < 0$ is defined to be the empty set.

Arguing iteratively for $k-1,k-2,...,0$ we find that
\begin{align*}
    \mathbb{E}  [  W'_{C,k}  & (a_Y)  W_{D,k}(a_Y,a_D)  Y_{k+1} (1-Y_{k}) (1-D_{k+1}) (1- C_{k+1}) \mid A=a_Y ] \\ 
        = \sum_{l} & \Big[ \Pr(Y_{k+1}=1 \mid Y_k= D_{k+1}=\bar{C}_{k+1}=0,L=l, A=a_Y) \\
         \prod_{j=0}^{k} &  \big\{ \Pr(D_{j+1}=0 \mid \bar{C}_{j+1}=D_{j}= Y_{j}=0, L = l,  A = a_Y)  \nonumber \\
      &  \times \Pr(Y_{j}=0 \mid \bar{C}_{j}=D_{j}= Y_{j-1}=0, L=l,  A = a_Y) \nonumber \\
     & \times \Pr(C_{j+1}= 0 \mid \bar{D}_j= \bar{Y}_{j}= \bar{C}_{j} =0,L = l,  a_Y) \big\}  \\
     \times    &  \Pr(L = l) W'_{C,k} (a_Y) W_{D,k}(a_Y,a_D)  \Big],  \\
\end{align*}

We plug in the expression for $ W'_{C,k} (a_Y) $ to get 
\begin{align*}
        =& \sum_{\bar{l}}  [ \Pr(Y_{k+1}=1 \mid Y_k= D_{k+1}=\bar{C}_{k+1}=0,L=l, A=a_Y) \\
        \times  & \prod_{j=0}^{k}  \big\{ \Pr(D_{j+1}=0 \mid \bar{C}_{j+1}=D_{j}= Y_{j}=0, L = l,  A = a_Y)  \nonumber \\
      &  \times \Pr(Y_{j}=0 \mid \bar{C}_{j}=D_{j}= Y_{j-1}=0, L = l,  A = a_Y) \big\} \nonumber \\
  & \times  \Pr(L = l)  W_{D,k}(a_Y,a_D) ],  \\
\end{align*}
We plug in the expression for the weights  $W_{D,k}(a_Y,a_D)$ to get 
\begin{align*}
        =& \sum_{\bar{l}}  [ \Pr(Y_{k+1}=1 \mid Y_k= D_{k+1}=\bar{C}_{k+1}=0,L=l, A=a_Y) \\
        \times  & \prod_{j=0}^{k}  \big\{ \Pr(D_{j+1}=0 \mid \bar{C}_{j+1}=D_{j}= Y_{j}=0, L = l,  A = a_D)  \nonumber \\
      &  \times \Pr(Y_{j}=0 \mid \bar{C}_{j}=D_{j}= Y_{j-1}=0, L = l,  A = a_Y) \big\},  \\
      \times & \Pr(L=l) ] \\
\end{align*}
and the final expression is equal to \eqref{eq: identification censoring L}.

\clearpage
\section{Exploring the dismissible component conditions}
\label{sec: exploring 1 and 2}
By considering causal graphs, we provide some insight into the interpretation of assumptions $\Delta$1 and $\Delta$2. 

\subsection{Scenario in which the dismissible component conditions are satisfied.}
Consider the study from Appendix \ref{sec: DCC as independences} in which $A_{Y}$ and $A_{D}$ were randomized without loss to follow-up, which ensures positivity and exchangeability. Furthermore, we assume that the usual assumptions about consistency is satisfied; if $A_{Y}=a_Y$,$A_{D}=a_D$, then $Y_k = {Y}^{ a_Y,  a_D}_k  $. \newline

%New Figure starts here
\begin{figure}
\centering
\begin{tikzpicture}
\begin{scope}[every node/.style={thick,draw=none}]
    \node (A) at (-1,0) {$A$};
    \node (Ay) at (1,1) {$A_{Y} \mid \color{red} a_Y$};
	\node (Ad) at (1,-1) {$A_{D} \mid \color{red} a_D$};
	\node (Y1) at (4,1) {$Y_1 \color{red} ^{a_Y,a_D} $};
    \node (D1) at (4,-1) {$D_1 \color{red} ^{a_Y,a_D} $};
    \node (Y2) at (7,1) {$Y_2 \color{red} ^{a_Y,a_D}  $};
    \node (D2) at (7,-1) {$D_2 \color{red} ^{a_Y,a_D}  $};
    \node (L) at (2,4) {$L$};
    \node (U) at (3,4) {$U_Y$};
    \node (W) at (3,-4) {$U_D$};
  %  \node (D3) at (8,-1) {$D(3)$};
\end{scope}

\begin{scope}[>={Stealth[black]},
              every node/.style={fill=white,circle},
              every edge/.style={draw=black,very thick}]
    \path [->] (A) edge[line width=0.85mm] (Ad);
    \path [->] (A) edge[line width=0.85mm] (Ay);
	\path [->] (Ad) edge (D1);
    \path [->] (Ad) edge[bend right] (D2);
	\path [->] (Ay) edge[bend left] (Y2);
    \path [->] (Ay) edge (Y1);	
    \path [->] (L) edge (Y1);
    \path [->] (L) edge (Y2);
    \path [->] (L) edge[bend right] (D1);
    \path [->] (L) edge (D2);
    \path [->] (Y1) edge (D2);
    \path [->] (Y1) edge (Y2);
    \path [->] (D1) edge (D2);
    \path [->] (D1) edge (Y1);
    \path [->] (D2) edge (Y2);
    \path [->] (W) edge (D1);
    \path [->] (W) edge (D2);
    \path [->] (U) edge (Y1);
    \path [->] (U) edge (Y2);
\end{scope}
\end{tikzpicture}
\caption{Single world intervention template (SWIT) that describes a scenario with interventions on $A_{Y}$, $A_{D}$ and $\bar{C}_k$. Even if $U_Y$ and $U_D$ are unmeasured, $\Delta$1 and $\Delta$2 hold.}
\label{fig: simple SWIG 2}
\end{figure}
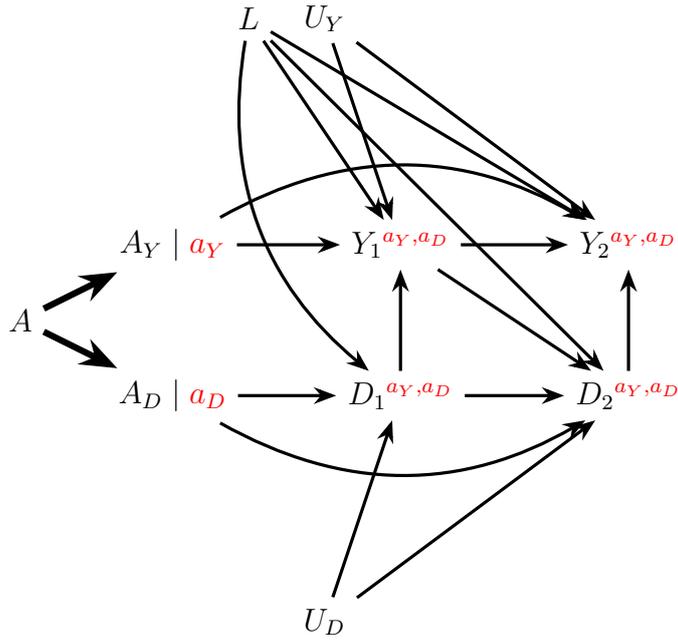

Assume that the causal structure in the  single world intervention template (SWIT) of Figure \ref{fig: simple SWIG} holds. Here, $A_{Y}$ is d-separated from both $Y^{ a_Y,a_D}_k$ and $D_k^{ a_Y,a_D}$ for $k \in 1,2$. Similarly $A_{D}$ is d-separated from both $Y^{ a_Y,a_D}_k$ and $D^{ a_Y,a_D}_k$. Hence, under the assumptions about positivity and consistency, we can identify the following joint law from the g-formula, 
\begin{align}
& \Pr(Y^{ a_Y,a_D}_2=1, Y^{ a_Y,a_D}_1=0, D^{ a_Y,a_D}_2=0,D^{ a_Y,a_D}_1=0 \mid L) \nonumber \\
=& \Pr(D_1=0 \mid A_{Y} = a_Y,A_{D} = a_D, L)\Pr(Y_1=0 \mid D_1=0, A_{Y} = a_Y,A_{D} = a_D,L) \nonumber \\
\times & \Pr(D_2=0 \mid D_1=0, Y_1=0, A_{Y} = a_Y,A_{D} = a_D,L) \nonumber \\
\times &\Pr(Y_2=1 \mid D_2=0, D_1=0, Y_1=0, A_{Y} = a_Y,A_{D} = a_D,L) \nonumber \\
=& \Pr(D_1=0 \mid A_{D} = a, L)\Pr(Y_1=0 \mid D_1=0, A_{Y} = a_Y,L) \nonumber \\
\times & \Pr(D_2=0 \mid D_1=0, Y_1=0, A_{D} = a_D,L) \Pr(Y_2=1 \mid D_2=0, D_1=0, Y_1=0, A_{Y} = a_Y,L), \nonumber \\
\end{align}
where the last equality follows due to conditional independences that we read off the causal graph. 
Similarly, we can identify 
\begin{align}
& \Pr(Y^{ a_Y,a_D}_1=0, D^{ a_Y,a_D}_2=0,D^{ a_Y,a_D}_1=0 \mid L) \nonumber \\
=& \Pr(D_1=0 \mid A_{D} = a_D, L)\Pr(Y_1=0 \mid D_1=0, A_{Y} = a_Y,L) \nonumber \\
\times & \Pr(D_2=0 \mid D_1=0, Y_1=0, A_{D} = a_D,L). \nonumber \\
\end{align}
Using laws of total probability, 
\begin{align}
& \Pr(Y^{ a_Y,a_D}_2=1 \mid Y^{ a_Y,a_D}_1=0, D^{ a_Y,a_D}_2=0,D^{ a_Y,a_D}_1=0, L) \nonumber \\
  = & \frac{\Pr(Y^{ a_Y,a_D}_2=1, Y^{ a_Y,a_D}_1=0, D^{ a_Y,a_D}_2=0,D^{ a_Y,a_D}_1=0 \mid L) }{\Pr(Y^{ a_Y,a_D}_1=0, D^{ a_Y,a_D}_2=0,D^{ a_Y,a_D}_1=0 \mid L)} \nonumber \\
   = & \Pr(Y_2=1 \mid D_2=0, D_1=0, Y_1=0, A_{Y} = a_Y,L). \nonumber \\
 \label{eq: id star nostar}
\end{align}
Hence, 
\begin{align*}
  =& \Pr(Y^{ a_Y,a_D=1}_2=1 \mid Y^{ a_Y,a_D=1}_1=0, D^{ a_Y,a_D=1}_2=0,D^{ a_Y,a_D=1}_1=0, L) \\
  =& \Pr(Y^{ a_Y,a_D=0}_2=1 \mid Y^{ a_Y,a_D=0}_1=0, D^{ a_Y,a_D=0}_2=0,D^{ a_Y,a_D=0}_1=0, L),
\end{align*}
that is  $\Delta$1 is satisfied at $k=2$. Using the same argument, we can derive that $\Delta$2 is satisfied for $k=2$, and both $\Delta$1 and $\Delta$2 will be satisfied for $k=1$. That is, Figure \ref{fig: simple SWIG} implies that $\Delta$1 and $\Delta$2 hold. Furthermore, we could use exactly the same derivations to find that $\Delta$1 and $\Delta$2 hold in Figure \ref{fig: simple SWIG 2}, even if $U_Y$ and $U_D$ are unmeasured. 

\clearpage
\subsection{Scenario in which the dismissible component conditions are not necessarily satisfied}
\begin{figure}
\centering
\begin{tikzpicture}
\begin{scope}[every node/.style={thick,draw=none}]
    \node (A) at (-1,0) {$A$};
    \node (Ay) at (1,1) {$A_{Y} \mid \color{red} a_Y$};
	\node (Ad) at (1,-1) {$A_{D} \mid \color{red} a_D$};
	\node (Y1) at (4,1) {$Y_1 \color{red} ^{a_Y,a_D} $};
    \node (D1) at (4,-1) {$D_1 \color{red} ^{a_Y,a_D} $};
    \node (Y2) at (7,1) {$Y_2 \color{red} ^{a_Y,a_D}  $};
    \node (D2) at (7,-1) {$D_2 \color{red} ^{a_Y,a_D}  $};
    \node (L) at (2.5,4) {$L$};
     \node (U) at (1.5,-3) {$U_{YD}$};
  %  \node (D3) at (8,-1) {$D(3)$};
\end{scope}

\begin{scope}[>={Stealth[black]},
              every node/.style={fill=white,circle},
              every edge/.style={draw=black,very thick}]
    \path [->] (A) edge[line width=0.85mm] (Ad);
    \path [->] (A) edge[line width=0.85mm] (Ay);
	\path [->] (Ad) edge (D1);
    \path [->] (Ad) edge[bend right] (D2);
	\path [->] (Ay) edge[bend left] (Y2);
    \path [->] (Ay) edge (Y1);	
    \path [->] (L) edge (Y1);
    \path [->] (L) edge (Y2);
    \path [->] (L) edge[bend right] (D1);
    \path [->] (L) edge (D2);
    \path [->] (Y1) edge (D2);
%    \path [->] (D1) edge (Y2);
    \path [->] (Y1) edge (Y2);
    \path [->] (D1) edge (D2);
    \path [->] (D1) edge (Y1);
    \path [->] (D2) edge (Y2);
   \path [->] (U) edge (Y1);
    \path [->] (U) edge (D1);
\end{scope}
\end{tikzpicture}
\caption{Single world intervention template (SWIT) of a scenario in which $\Delta$1 and $\Delta$2 are not implied by the graph.}
\label{fig: simple SWIG violation}
\end{figure}
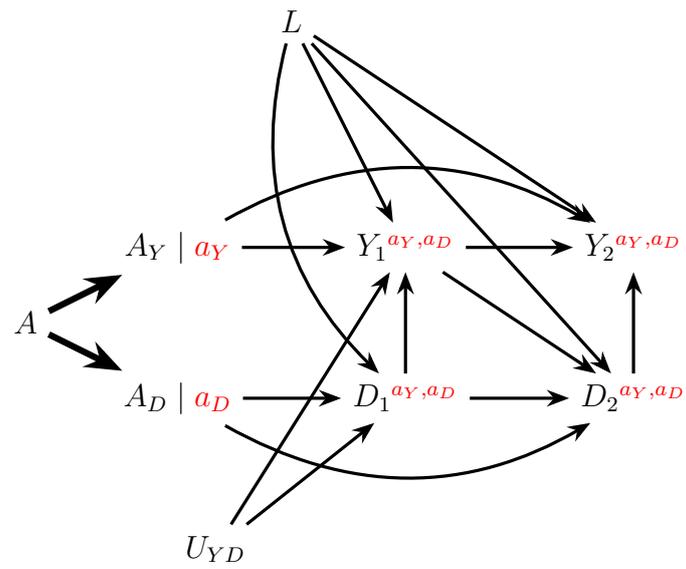

Consider the SWIT in Figure \ref{fig: simple SWIG violation}, which only differs from Figure \ref{fig: simple SWIG} in the variable $U_Y$ that is an unmeasured common cause of $Y_1$ and $D_1$. Here we read off Figure \ref{fig: simple SWIG violation} to find that 
\begin{align}
& \Pr(Y^{ a_Y,a_D}_1=1 \mid D^{ a_Y,a_D}_1=0, L) \nonumber \\
=& \Pr(Y_1=1 \mid  D_1=0, A_{Y}= a_Y, A_{D}= a_D,L), \nonumber \\
\end{align}

However, we cannot conclude from the graph that 
\begin{align}
     & \Pr(Y_1=1 \mid  D_1=0,  A_{Y} = a_Y, A_{D} = 1,L) \nonumber \\
     =& \Pr(Y_1=1 \mid  D_1=0, A_{Y} = a_Y, A = 0,L) \nonumber \\
\end{align}
because there is an open collider path $a_D \rightarrow D_1 \leftarrow U_{YD} \rightarrow Y_1$. Hence, we cannot conclude that the graph in Figure \ref{fig: simple SWIG violation} implies $\Delta$1, and our results do not allow us to identify $ \Pr(Y^{ a_Y,a_D}_1=1)$ in this scenario. The unmeasured common cause $U_{YD}$ of $Y_k$ and $D_{k'}$ for $k,k' \in (0,1,...,K+1)$ leads to violation of $\Delta$1 and $\Delta$2.  

\clearpage
\section{Simulations}
\label{sec app: simulations}
Here we present simulations from 5 scenarios to illustrate the finite sample performance of the separable effects. We consider settings where the dismissible component conditions are satisfied, but also settings where these conditions are violated. Furthermore, we consider coverage under violation of the parametric model assumptions. % the separable effects may be different from the total effect, ii) that interpretation and estimation of the separable effects crucially depend on the dismissible component conditions and iii) that the estimators seem to have adequate finite sample behaviour. 

In each scenario, we simulated two randomized experiments in which 400 and 2000 subjects were randomly assigned to treatment $A \in \{0,1\}$, respectively. To assess finite sample behavior, we calculated confidence intervals for 3 time points by simulating each experiment 500 times, and for each of these experiments we created non-parametric percentile bootstrap confidence intervals from 500 bootstrap samples. 

The true cumulative incidences from the simulation scenarios are shown in Figure \ref{fig:simulation setups}. Generally, our simulations confirm that the g-formula and IPW estimators perform satisfactory when the identifiability conditions are satisfied.

\subsection{Data generating mechanism}
For each individual, the data were generated from the following algorithm, where we have omitted $i$ subscripts to indicate inidivuals:
\begin{enumerate}
    \item Draw $L_1 \sim \text{Bernoulli} [p=0.25]$.
    \item Draw $L_2 \sim \text{Bernoulli}[p=0.2L_1+0.8(1-L_1)]$.
    \item Draw $A  \sim \text{Bernoulli} [p=0.5]$, and define $A_Y \equiv A_D \equiv A$.
    \item Set $D_0=Y_0=0$. 
    \item For each $k \in \{0,K\}$, 
    \begin{itemize}
        \item if $D_k=Y_k=0$, \\
    draw $D_{k+1} \sim \text{Bernoulli} [p=\alpha_D \psi_k(A_Y,A_D,L_1,L_2)]$, where
    \begin{align}
    \psi_k(A_Y,A_D,L_1,L_2) = \text{expit} & (\omega_{0} + \omega_{1,k}k  + \omega_{2}A_Y + \omega_{3}A_D + \omega_{4}L_1+\omega_{5}L_2  \nonumber \\
     & +  \omega_{6}A_YL_1 + \omega_{7}A_DL_1  )
    \end{align} \\
    if $D_{k+1}=0$, \\
    draw $Y_{k+1} \sim \text{Bernoulli} (p=\alpha_Y \lambda_k(A_Y,A_D,L_1,L_2))$, where
    \begin{align}
    \lambda_k(A_Y,A_D,L_1,L_2) = \text{expit} & (\xi_{0} + \xi_{1,k}k  + \xi_{2}A_Y + \xi_{3}A_D + \xi_{4}L_1 + \xi_{5}L_2  \nonumber \\
     & +  \xi_{6}A_YL_1 + \xi_{7}A_DL_1   ).
    \end{align} \\
    if $D_{k+1}=1$, set $Y_{k+1}=0$.
    \item else, define $D_{k+1}=D_{k}$,$Y_{k+1}=Y_{k}$. 
    \end{itemize}
\end{enumerate}

The coefficients in each of the scenarios are found in Table \ref{tab: sim coef} and the true cumulative incidence curves of $Y_{k+1}, k \in \{0,99\}$ is found in Figure \ref{fig:simulation setups}. % $\omega_j, \xi_j$, where $j \in \{0,9\}$

%\caption{Simulation settings.}
%\label{tab: sim coef}
% latex table generated in R 3.5.1 by xtable 1.8-3 package
% Tue Oct  1 10:24:09 2019
\begin{table}[ht]
\centering
\begin{tabular}{rrrrrrrrrrrrrrrrrrr}
  \hline
Scenario & $\alpha_Y$ & $\xi_1$ & $\xi_2$ & $\xi_3$ & $\xi_4$ & $\xi_5$ & $\xi_6$ & $\xi_7$  & $\alpha_D$ & $\omega_1$ & $\omega_2$ & $\omega_3$ & $\omega_4$ & $\omega_5$ & $\omega_6$ & $\omega_7$  \\ 
  \hline
   1 & 0.01 & 0 & 10 & 0 & 5 & 0 & 0 & 0 & 0.03 &  0 & 0 & -2 & 5 & 0 & 0 & 0   \\ 
   2 & 0.01 & 0 & 10 & 0 & -2 & 5 & 0 & 0  & 0.03  & 0 & 0 & -2 & 5 & -2 & 0 & 0  \\ 
   3 & 0.01 & 0 & 10 & 0 & 5 & -10 & 5 & 0  & 0.03 & 0 & 0 & -2 & 5 & -10 & 0 & 0  \\ 
   4 & 0.01 & 0 & 10 & 5 & 5 & 0 & 0 & 0 &  0.03 & 0 & 0 & -2 & 5 & 0 & 0 & 0   \\ 
   5 & 0.01 & 0 & 10 & 0 & -10 & 0 & 0 & 0 & 0.03 & 0 & 0 & -2 & 0 & 0 & 0 & 0 \\ 
   \hline
\end{tabular}
\caption{Data generating mechanism for the 7 simulation scenarios.}
\label{tab: sim coef}
\end{table}

\clearpage

\subsection{Scenario 1: Dismissible component conditions hold and no model mis-specification.}
Data were generated from the simple setting described by the first row in Table \ref{tab: sim coef}; that is, there is a causal effect of (i) $A_Y$ on $Y_k$, (ii) $A_D$ on $D_k$, and (iii)  $L_1$ on both $Y_k$ and $D_k$. Here, both the dismissible component conditions hold conditional on $L_1$. 

To estimate the separable effects, we fitted the following models
\begin{align}
    &  \text{logit} [\widehat{\Pr}(Y_{k}=1 \mid D_{k}= Y_{k-1} = 0,A, L_1, L_2)]  = \theta_{0,k} + \theta_{1}A + \theta_{2}L_1 + \theta_{3}L_2 \label{eq: sim estim mod y}  \\
& \text{logit} [\widehat{\Pr}(D_{k}=1 \mid D_{k-1}= Y_{k-1}=0,A, L_1, L_2)]  = \beta_{0,k} + \beta_{1}A + \beta_{2}L_1, \label{eq: sim estim mod d}
\end{align}

which are correctly specified, even if model \eqref{eq: sim estim mod y} includes a term $\theta_{3}$ that is redundant. Thus, we would expect all our estimators to have nominal coverage, and this is confirmed in Table \ref{tab: setting 1}; here, coverage is derived from estimated 95\% confidence intervals based on the parametric g-formula estimator (g-formula) and the weighted estimators ($\hat{\nu}_{1,a_Y,a_D,k}$ and $\hat{\nu}_{2,a_Y,a_D,k}$) for the trial with $n=400$ subjects. %; all the estimators perform satisfactory. This is expected because all the parametric assumptions are satisfied.

\begin{table}[ht]
\centering
\begin{tabular}{lr|rrr|}
 & & \multicolumn{3}{c|}{$n=400$} \\
 Parameter & Estimator  & $k=100$ & $k=75$ & $k=25$ \\ 
\hline
$ \Pr(Y^{a_Y=1,a_D=1}_{k}=1)$ & g-formula &  0.95 & 0.94 & 0.93 \\ 
   &  non-parametric & 0.95 & 0.94 & 0.95 \\ 
   \hline
$ \Pr(Y^{a_Y=0,a_D=0}_{k}=1)$ & g-formula & 0.94 & 0.93 & 0.92 \\ 
  &  non-parametric & 0.94 & 0.95 & 0.95 \\ 
   \hline
$ \Pr(Y^{a_Y=1,a_D=0}_{k}=1)$ & g-formula & 0.95 & 0.96 & 0.94 \\ 
   & $\hat{\nu}_{1,a_Y,a_D,k} $ & 0.94 & 0.95 & 0.95 \\ 
  & $\hat{\nu}_{2,a_Y,a_D,k} $ & 0.96 & 0.95 & 0.95 \\ 
   \hline
$ \Pr(Y^{a_Y=0,a_D=1}_{k}=1)$ & g-formula & 0.93 & 0.93 & 0.94 \\ 
   & $\hat{\nu}_{1,a_Y,a_D,k} $ & 0.92 & 0.90 & 0.95 \\ 
 & $\hat{\nu}_{2,a_Y,a_D,k} $ & 0.94 & 0.94 & 0.92 \\
  \end{tabular}
  \caption{Scenario 1.}
  \label{tab: setting 1}
\end{table}

\clearpage

\subsection*{Scenario 2: Dismissible component conditions hold and minor model mis-specification.}
In this scenario, there are causal effects of both $L_1$ and $L_2$ on $Y_k$ and $D_k$ (second row in Table \ref{tab: sim coef}). Both the dismissible component conditions hold conditional on $L_1$ and $L_2$. We used regression models \eqref{eq: sim estim mod y} and \eqref{eq: sim estim mod d} for model fitting. 

\begin{figure}
\includegraphics[scale=1.0]{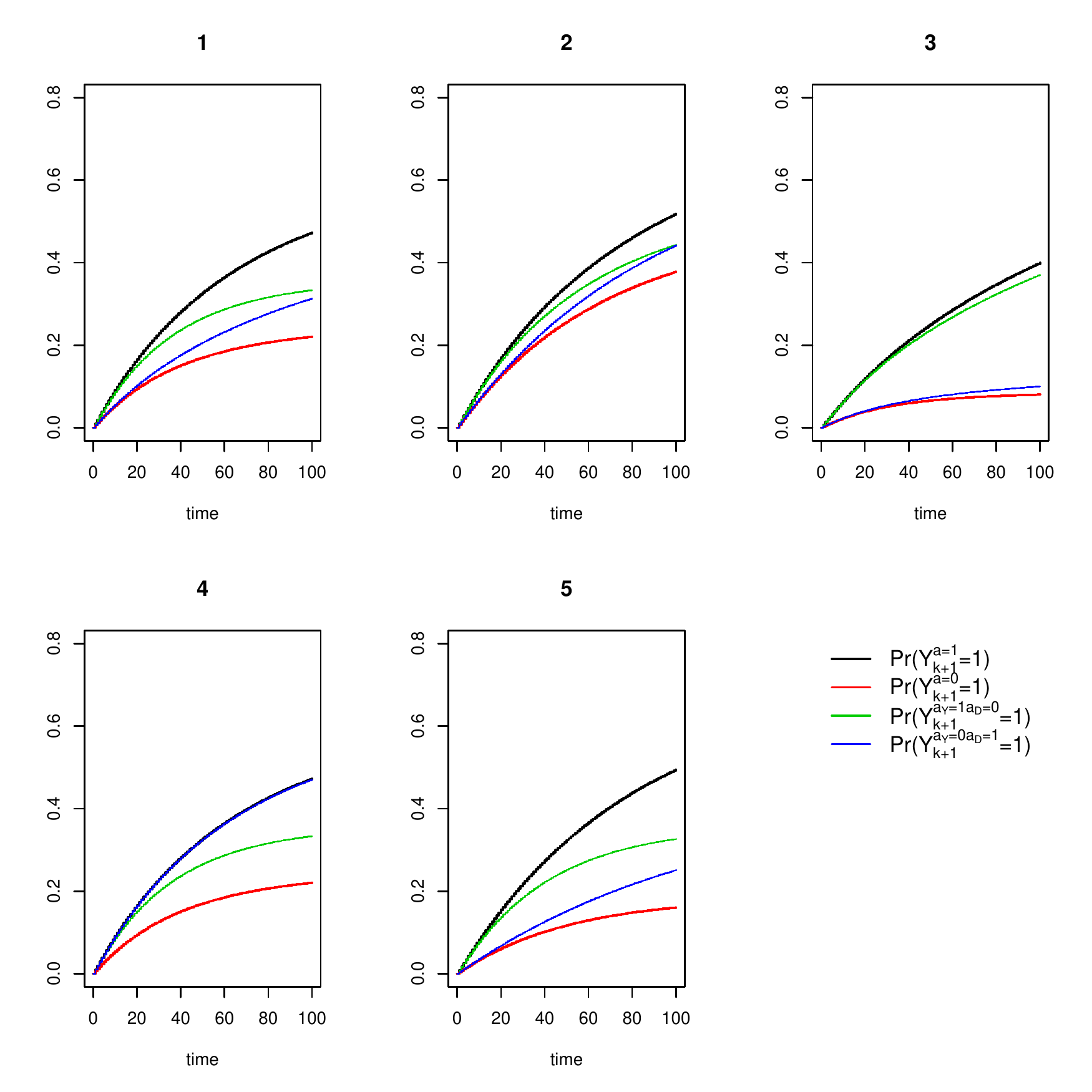}
\caption{True cumulative incidence curves for scenarios 1-5.}
\label{fig:simulation setups}
\end{figure}

Note that in this setting \eqref{eq: sim estim mod y} is correctly specified, but \eqref{eq: sim estim mod d} is mis-specified because it does not include a term for $L_2$. Thus, we would expect that the IPW estimator that uses the correctly specified regression model ( $\hat{\nu}_{2,a_Y,a_D,k} $) is unbiased, but the parametric g-formula estimator and the other IPW estimator ( $\hat{\nu}_{1,a_Y,a_D,k} $) are biased because \eqref{eq: sim estim mod d} is mis-specified. The results in Table \ref{tab: setting 2}, however, suggest that all estimators have close to nominal coverage. This may be explained by the fact that the model mis-specification is minor, and the magnitude of the separable effects is small (see Figure \ref{fig:simulation setups}). 

\begin{table}[ht]
\centering
\begin{tabular}{lr|rrr|}
 & & \multicolumn{3}{c|}{$n=400$} \\
 Parameter & Estimator  & $k=100$ & $k=75$ & $k=25$ \\ 
\hline
$ \Pr(Y^{a_Y=1,a_D=1}_{k}=1)$ & g-formula &  0.91 & 0.92 & 0.91 \\ 
&  non-parametric  & 0.95 & 0.96 & 0.93 \\ 
   \hline
$ \Pr(Y^{a_Y=0,a_D=0}_{k}=1)$ & g-formula &  0.94 & 0.94 & 0.93 \\ 
  &  non-parametric  & 0.93 & 0.93 & 0.93 \\ 
   \hline
$ \Pr(Y^{a_Y=1,a_D=0}_{k}=1)$ & g-formula & 0.96 & 0.94 & 0.91 \\ 
& $\hat{\nu}_{1,a_Y,a_D,k} $ & 0.93 & 0.95 & 0.93 \\ 
& $\hat{\nu}_{2,a_Y,a_D,k} $ &  0.91 & 0.92 & 0.88 \\ 
   \hline
$ \Pr(Y^{a_Y=0,a_D=1}_{k}=1)$ & g-formula &  0.94 & 0.93 & 0.93 \\ 
& $\hat{\nu}_{1,a_Y,a_D,k} $ &  0.90 & 0.91 & 0.93 \\ 
& $\hat{\nu}_{2,a_Y,a_D,k} $ & 0.93 & 0.94 & 0.94 \\ 
  \end{tabular}
    \caption{Scenario 2.}
  \label{tab: setting 2}
\end{table}

\clearpage 

\subsection*{Scenario 3: Dismissible component conditions hold and model mis-specification}
% latex table generated in R 3.5.1 by xtable 1.8-3 package
% Tue Oct  1 16:17:31 2019
In this scenario, both the dismissible component conditions hold conditional on $L_1$ and $L_2$. Unlike Scenarios 1 and 2, we fitted the following regression models to the simulated data,
\begin{align}
    &  \text{logit} [\widehat{\Pr}(Y_{k}=1 \mid D_{k}= Y_{k-1} = 0,A, L_1, L_2)] = \theta_{0,k} + \theta_{1}A + \theta_{2}L_1  \label{eq: sim estim mod y 2}  \\
& \text{logit} [\widehat{\Pr}(D_{k}=1 \mid D_{k-1}= Y_{k-1}=0,A, L_1, L_2)] = \beta_{0,k} + \beta_{1}A + \beta_{2}L_1 + \beta_{3}L_2 \label{eq: sim estim mod d 2}.
\end{align}

Here, \eqref{eq: sim estim mod y 2} is mis-specified because it does not include a term for $L_2$, but \eqref{eq: sim estim mod d 2} is correctly specified; thus the correctness of the model specifications are opposite from Scenario 2. Also, $L_2$ exerts larger effects on $Y_k$ and $D_k$ in this setting compared to Scenario 2.

The results in Table \ref{tab: setting 3} illustrate that the IPW estimator $\hat{\nu}_{1,a_Y,a_D,k} $ is unbiased because it relies on a correctly specified model, but the parametric g-formula estimator and the other IPW estimator ($\hat{\nu}_{2,a_Y,a_D,k} $) are biased -- in particular, for shorter follow-up times -- because they rely on mis-specified regression models. 

\begin{table}[ht]
\centering
\begin{tabular}{lr|rrr|}
 & & \multicolumn{3}{c|}{$n=400$} \\
 Parameter & Estimator  & $k=100$ & $k=75$ & $k=25$ \\ 
\hline
$ \Pr(Y^{a_Y=1,a_D=1}_{k}=1)$ & g-formula & 0.93 & 0.95 & 0.91 \\ 
  &  non-parametric & 0.93 & 0.93 & 0.94 \\ 
   \hline
$ \Pr(Y^{a_Y=0,a_D=0}_{k}=1)$ & g-formula & 0.93 & 0.86 & 0.48 \\ 
 &  non-parametric & 0.94 & 0.93 & 0.94 \\ 
   \hline
$ \Pr(Y^{a_Y=1,a_D=0}_{k}=1)$ & g-formula & 0.93 & 0.94 & 0.93 \\ 
 & $\hat{\nu}_{1,a_Y,a_D,k} $ & 0.94 & 0.94 & 0.93 \\ 
 & $\hat{\nu}_{2,a_Y,a_D,k} $ & 0.91 & 0.72 & 0.56 \\ 
   \hline
$ \Pr(Y^{a_Y=0,a_D=1}_{k}=1)$ & g-formula & 0.82 & 0.74 & 0.45 \\ 
 & $\hat{\nu}_{1,a_Y,a_D,k} $ & 0.95 & 0.95 & 0.94 \\ 
 & $\hat{\nu}_{2,a_Y,a_D,k} $ & 0.84 & 0.72 & 0.33 \\ 
  \end{tabular}
      \caption{Scenario 3.}
  \label{tab: setting 3}
\end{table}
\clearpage

\subsection*{Scenario 4: Dismissible component conditions fail and model misspecification.}
The dismissible component condition $\Delta 2$ fails in this scenario due to the non-zero coefficient $\omega_3 = 5$; there is a direct effect $A_Y \rightarrow D_k$ for $k \in \{0,100\}$. Yet we fitted regression models \eqref{eq: sim estim mod y} and \eqref{eq: sim estim mod d} to the simulated data. %, which means that \eqref{eq: sim estim mod y} is mis-specified. % \eqref{eq: sim estim mod d} is correctly specified. 

The simulations suggest that none of the estimators has nominal coverage for $\Pr(Y^{a_Y=0,a_D=1}_{k+1}=1)$. However, since dismissible component condition $\Delta 1$ holds we can identify $ \Pr(Y^{a_Y=1,a_D=0}_{k+1}=1)$, as suggested by the nominal coverage for this quantity in Table \ref{tab: setting 4}. Yet we cannot interpret a contrast $ \Pr(Y^{a_Y=0,a_D=1}_{k+1}=1) \text{ vs } \Pr(Y^{a_Y=1,a_D=1}_{k+1}=1) $ as the \textit{separable direct effect} of $A$, due to the violation of the dismissible component condition.

\begin{table}[ht]
\centering
\begin{tabular}{lr|rrr|}
 & & \multicolumn{3}{c|}{$n=400$} \\
 Parameter & Estimator  & $k=100$ & $k=75$ & $k=25$ \\ 
\hline
$ \Pr(Y^{a_Y=1,a_D=1}_{k}=1)$ & g-formula & 0.96 & 0.94 & 0.93 \\ 
    &  non-parametric & 0.95 & 0.94 & 0.94 \\ 
   \hline
 $ \Pr(Y^{a_Y=0,a_D=0}_{k}=1)$ & g-formula & 0.93 & 0.93 & 0.92 \\ 
  &  non-parametric & 0.93 & 0.93 & 0.95 \\ 
   \hline
 $ \Pr(Y^{a_Y=1,a_D=0}_{k}=1)$ & g-formula  & 0.96 & 0.97 & 0.94 \\ 
& $\hat{\nu}_{1,a_Y,a_D,k} $ & 0.94 & 0.96 & 0.94 \\ 
& $\hat{\nu}_{2,a_Y,a_D,k} $ & 0.97 & 0.96 & 0.96 \\ 
   \hline
 $ \Pr(Y^{a_Y=0,a_D=1}_{k}=1)$ & g-formula  &  0.05 & 0.05 & 0.07 \\ 
& $\hat{\nu}_{1,a_Y,a_D,k} $ & 0.31 & 0.26 & 0.34 \\ 
& $\hat{\nu}_{2,a_Y,a_D,k} $ & 0.05 & 0.04 & 0.12 \\
  \end{tabular}
      \caption{Scenario 4.}
  \label{tab: setting 4}
\end{table}

\clearpage

\subsection*{Scenario 5: Dismissible component conditions hold and no model misspecification.}
In this scenario, $L_1$ exerts (strong) causal effects on $Y_k$ but not on $D_k$. Thus, all the dismissible component conditions hold marginally. To illustrate that we obtain unbiased estimates even if $L_1$ is not included in any of the regression models, we fitted the parsimonious models,
\begin{align}
    &  \text{logit} [\widehat{\Pr}(Y_{k}=1 \mid D_{k}= Y_{k-1} = 0,A)] = \theta_{0,k} + \theta_{1}A.  \label{eq: sim estim mod y 3}  \\
& \text{logit} [\widehat{\Pr}(D_{k}=1 \mid D_{k-1}= Y_{k-1}=0,A)] = \beta_{0,k} + \beta_{1}A, \label{eq: sim estim mod d 3}
\end{align}
and the results in Table \ref{tab: setting 5} show that all estimators have nominal coverage, even if $L_1$ is not included in the models.

% latex table generated in R 3.5.1 by xtable 1.8-3 package
% Tue Oct  1 16:18:49 2019
\begin{table}[ht]
\centering
\begin{tabular}{lr|rrr|}
 & & \multicolumn{3}{c|}{$n=400$} \\
 Parameter & Estimator  & $k=100$ & $k=75$ & $k=25$ \\ 
\hline
$ \Pr(Y^{a_Y=1,a_D=1}_{k}=1)$ & g-formula & 0.95 & 0.94 & 0.94 \\ 
    &  non-parametric & 0.95 & 0.95 & 0.95 \\ 
   \hline
 $ \Pr(Y^{a_Y=0,a_D=0}_{k}=1)$ & g-formula & 0.94 & 0.94 & 0.93 \\ 
    &  non-parametric & 0.95 & 0.94 & 0.94 \\ 
   \hline
 $ \Pr(Y^{a_Y=1,a_D=0}_{k}=1)$ & g-formula  & 0.96 & 0.95 & 0.94 \\ 
   & $\hat{\nu}_{1,a_Y,a_D,k} $ & 0.97 & 0.96 & 0.95 \\ 
    & $\hat{\nu}_{2,a_Y,a_D,k} $ & 0.95 & 0.95 & 0.94 \\ 
   \hline
 $ \Pr(Y^{a_Y=0,a_D=1}_{k}=1)$ & g-formula & 0.93 & 0.94 & 0.94 \\ 
    &  $\hat{\nu}_{1,a_Y,a_D,k} $ & 0.94 & 0.93 & 0.94 \\ 
    & $\hat{\nu}_{2,a_Y,a_D,k} $ & 0.94 & 0.94 & 0.94 \\ 
\end{tabular}
      \caption{Scenario 5.}
  \label{tab: setting 5}
\end{table}

\clearpage

\end{document}